\documentclass[a4]{paper}
\usepackage[dvips]{graphics}
\usepackage[pdftex]{graphicx}
\usepackage{psfrag}
\usepackage{fancyhdr}
\usepackage{subfigure}
\usepackage{layout}
\usepackage{mathpple}
\usepackage{color}
\usepackage{amssymb,amsmath}
\usepackage{texdraw} 
\usepackage{color}

\title{Corner-Impact Bifurcations: a novel class of discontinuity-induced bifurcations in Cam-Follower Systems
\thanks{This work was partially supported by the European Project SICONOS IST2001-37172.}}

\author{Gustavo Osorio$^{\ddag}$\thanks{Corresponding Author. Departamento de Ingenier\'ia El\'ectrica, Electr\'onica
        y Computaci\'on, Universidad Nacional de Colombia, Carrera 26 \#64-60,
        Manizales, Colombia ({\tt gosorio@unina.it, gaosoriol@unal.edu.co}).} $^{\ddag}$
        \and Mario di Bernardo\thanks{Dipartimento di Informatica e
        Sistemistica, Universit\'a degli Studi di Napoli Federico II,
        Via Claudio 21, 80125, Napoli, Italia ({\tt
        mario.dibernardo@unina.it,gosorio@unina.it,stefania.santini@unina.it}).}
        \and Stefania Santini$^{\ddag}$}

\begin{document}

\maketitle

\begin{abstract}
This paper is concerned with the analysis of a class of impacting
systems of relevance in applications: cam-follower systems. We
show that these systems, which can be modelled as discontinuously
forced impact oscillators, can exhibit complex behaviour due to
the detachment at high rotational speeds between the follower and
the cam. We propose that the observed phenomena can be explained
in terms of a novel type of discontinuity-induced bifurcation,
termed as corner-impact. We present a complete analysis of this
bifurcation in the case of non-autonomous impact oscillator and
explain the transition to chaos observed in a representative
cam-follower example. The theoretical findings are validated
numerically.
\end{abstract}



\section{Introduction}
Recently, much research effort has been spent to analyse the
dynamics of piecewise smooth dynamical systems with impacts
\cite{Br:99,ZhMo:03}. These systems arise in many areas of
engineering and applied science. A typical example is that of
mechanical systems characterised by structural components with
displacement constraints. Examples include bouncing or hopping
robots, systems with backlash or friction, gears, vibro-impacting
mechanical devices \cite{Br:99}.

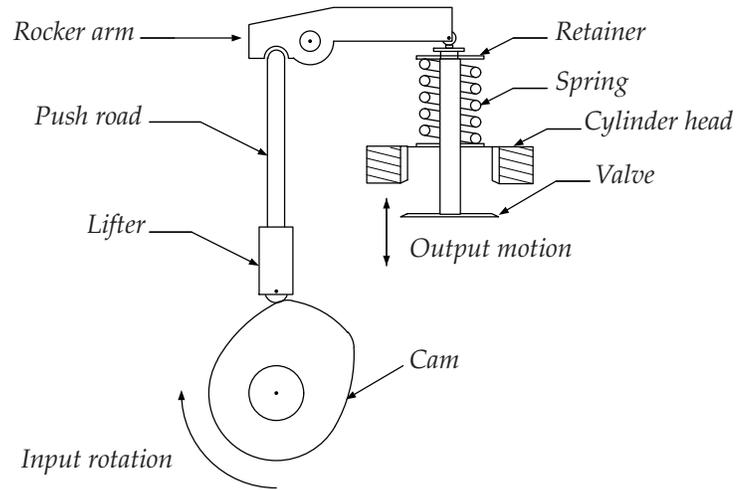
\begin{figure}
\begin{center}
\begin{picture}(150,180)(0,0)
\begin{texdraw}
\setunitscale 0.45
 \drawdim cm \linewd 0.02 \arrowheadtype t:F
\arrowheadsize l:0.16 w:0.08
 \move(0 0) 
 \linewd 0.03
 \setgray 0
 \move(5.25 3.0) \larc r:2 sd:150.1794 ed:330.1794 
 \larc r:2.8 sd:180 ed:270 \rmove(-2.8 0) \arrowheadtype t:F \arrowheadsize l:0.3 w:0.2  \ravec(0 0.1) 
 \move(5.25 3.0) \larc r:2.77 sd: 40.1794 ed:80.1794 
 \move(7.2021    1.8811) \larc r:4.25 sd:117.3427 ed:150.1794 
 \move(3.2979    4.1189) \larc r:4.25 sd:-29.8206 ed:3.0161 
 \move(5.5945    4.9901) \larc r:0.745 sd:80.1794 ed:117.3427 
 \move(6.7931    4.3031) \larc r:0.745 sd: 3.0161 ed:40.1794 
 \move(5.25 3.0) \lcir r:0.8  
 \move(5.25 3.0) \fcir f:0 r:0.05 
 \move(8.5 6.75) \ravec(0 2) \ravec(0 -2)
 \move(14.5 9.25) \rlvec(-1 0) \ravec(-1.7 -1)
 \move(13.25 12) \rlvec(-1 0) \ravec(-1 -0.5)
 \move(14.25 10.8) \rlvec(-1 0) \ravec(-1 -0.5)
 \move(13.25 13.5) \rlvec(-1 0) \ravec(-1 -0.5)
 \move(9 3.8) \ravec(-1.7 -1)
 \move(1.5 7.75) \rlvec(1.25 0) \ravec(2 -1)
 \move(1.5 11) \rlvec(1.5 0) \ravec(2 -1)
 \move(1.25 13.5) \rlvec(1 0) \ravec(2 0)
 \move(5.25 6) \larc r:0.335 sd:200 ed:340 
 \fcir f:0 r:0.05
 \move(5.25 6) \rmove(-0.5 -0.1) \rlvec(1 0) \rlvec(0 2) \rlvec(-1 0) \rlvec(0 -2)
 \move(5.25 8) \rmove(0.25 -0.1) \rlvec(0 5) \rmove(-0.5 0) \rlvec(0 -5)
 \move(5.25 10) \rmove(0 2.9) \larc r:0.25 sd: 0 ed:180 \larc r:0.35 sd: 0 ed:180
 \move(5.25 10) \rmove(0.35 2.9) \rlvec(0.15 0)
 \move(5.25 10) \rmove(1 3.4) \larc r:0.7 sd: 225 ed:360 \lcir r:0.3 \fcir f:0 r:0.05
 \move(5.25 10) \rmove(1.7 3.4) \rlvec(3.5 0) \rlvec(0 1) \rlvec(-3.5 0) \rlvec(-2.5
 -0.50) \rlvec(0 -1) \rlvec(0.45 0)
 \move(5.25 10) \rmove(5.1 3.5) \larc r:0.2 sd:210  ed:60 \fcir f:0 r:0.05
 \move(5.25 10) \rmove(5 3.3) \rlvec(0.2 0)\rlvec(0 -0.1)\rlvec(-0.2 0) \rlvec(0 0.1)
 \rmove(-0.35 -0.1) \rlvec(0.9 0) \rlvec(0 -0.1)\rlvec(-0.9 0) \rlvec(0 0.1)
 \rmove(0.2 -0.1) 
 \rlvec(0 -4.8) \rlvec(0.5 0) \rlvec(0 4.8)
 \rmove(-1.5 -4.8) \rlvec(2.5 0) \rlvec(0.2 -0.1) \rlvec(-2.9 0)
 \rlvec(0.2 0.1)
 \rmove(1.25 0) \rmove(-1.25 1) \rlvec(-0.2 -0.1) \rlvec(0 1.1)
 \rlvec(0.2 0) \rlvec(0 -1)
 \rmove(-0.2 -0.1) \rlvec(-1 0) \rlvec(0 1.1) \rlvec(1 0)
 \linewd 0.01
 \rmove(0 -1.1) \rlvec(-1 0.25)
 \rmove(1 0.0) \rlvec(-1 0.25)
 \rmove(1 0.0) \rlvec(-1 0.25)
 \rmove(1 0.0) \rlvec(-1 0.25)
 \rmove(1 0.0) \rlvec(-0.3 0.075)
 \rmove(0.5 0.0125)
 \linewd 0.03
 \rlvec(3.7 0) \rlvec(0 -1.1) \rlvec(-1 0) \rlvec(-0.2 0.1)
 \rlvec(0 1) \rmove(0.2 0) \rlvec(0 -1.1)
 \linewd 0.01
 \rmove(0 0.25) \rlvec(1 -0.25)
 \rmove(-1 0.5) \rlvec(1 -0.25)
 \rmove(-1 0.50) \rlvec(1 -0.25)
 \rmove(-1 0.50) \rlvec(1 -0.25)
 \rmove(-0.44 0.3) \rlvec(0.44 -0.11)
 \rmove(-1.7 0.11)
 \linewd 0.03
 \rmove(0.25 0.1) \rlvec(-2 0) \rlvec(0 -0.1) \rmove(2 0)
 \rlvec(0 0.1) \rmove (-0.25 -0.1)
 \rmove(0 0.25)
 \lcir r:0.15 \rmove(0 -0.15) \rlvec(-1.5 0.25) \rmove(1.5 0.05) \rlvec(-1.5 0.25) \rmove(0 -0.15)
 \lcir r:0.15 \rmove(1.5 0.25)
 \lcir r:0.15 \rmove(0 -0.15) \rlvec(-1.5 0.25) \rmove(1.5 0.05) \rlvec(-1.5 0.25) \rmove(0 -0.15)
 \lcir r:0.15 \rmove(1.5 0.25)
 \lcir r:0.15 \rmove(0 -0.15) \rlvec(-1.5 0.25) \rmove(1.5 0.05) \rlvec(-1.5 0.25) \rmove(0 -0.15)
 \lcir r:0.15 \rmove(1.5 0.25)
 \lcir r:0.15 \rmove(0 -0.15) \rlvec(-1.5 0.25) \rmove(1.5 0.05) \rlvec(-1.5 0.25) \rmove(0 -0.15)
 \lcir r:0.15 \rmove(1.5 0.25)
 \lcir r:0.15 \rmove(0 -0.15) \rlvec(-1.5 0.25) \rmove(1.5 0.05) \rlvec(-1.5 0.25) \rmove(0 -0.15)
 \lcir r:0.15 \rmove(-0.25 0.15) \rlvec(2 0) \rlvec(0 0.1)
 \rlvec(-2 0) \rlvec(0 -0.1)
 \rmove(0.7 0) \rlvec(0.6 0) \rlvec(0 -4.6) \rlvec(-0.6 0) \rlvec(0 4.6) \lfill
 f:1
 \linewd 0.02
\end{texdraw}
 \put (-215,10){\mbox{\textit{Input rotation}}}
 \put (-68,48){\mbox{\textit{Cam}}}
 \put (-68,90){\mbox{\textit{Output motion}}}
 \put (2,118){\mbox{\textit{Valve}}}
 \put (-2,138){\mbox{\textit{Cylinder head}}}
 \put (-13,153.0){\mbox{\textit{Spring}}}
 \put (-13,172){\mbox{\textit{Retainer}}}
 \put (-190,99){\mbox{\textit{Lifter}}}
 \put (-210,140){\mbox{\textit{Push road}}}
 \put (-218,172){\mbox{\textit{Rocker arm}}}
\end{picture}
\end{center}
\caption{Valve train configuration.}
\label{fig:pushroadvalve}
\end{figure}

Cam-follower systems are a particularly important class of
mechanical systems with displacement constraints widely used for
the operation of various machines and mechanical devices
\cite{Norton:02}. Usually, their purpose is to actuate valves or
other mechanisms through the movement of a follower forced by a
rotating cam. For example, all types of automated production
machines, including screw machines, spring winders and assembly
machines, rely heavily on this kind of systems for their
operation.  One of the most common application is to the valve
train of internal combustion engines (ICE) \cite{Heywood}, where
the effectiveness of the ICE is based on the proper working of a
cam-follower system.  A schematic of a single valve for a typical
pushrod type engine is presented in Figure
\ref{fig:pushroadvalve}. Here, the cam rotation results in a
linear motion imparted to the valve. The valve spring in the
system provides the restoring force necessary to maintain contact
between the components.

To guarantee that the follower moves as required, it is important
in applications to carefully design the cam profile. Different cam
geometries are used in practice ranging from circular cams to
highly complex cam profiles. In general, there is now a large
variety of alternative methods to select the cam profile. For
example, by using constrained optimization algorithm, it is
possible to use splines to obtain the cam geometry from the
desired motion that the cam is required to impart on the follower
(see for examples \cite{cardona} and \cite{fabien}). This often
means that while the cam has a continuous displacement profile, it
might have discontinuities in its acceleration \cite{nortonsae}.

It has been observed that, as the cam rotational speed increases,
the follower can detach from the cam. This causes the onset of
 undesired behaviour associated to impacts taking place
between the follower and the cam. For example, in automotive
engines this phenomenon can seriously deteriorate the engine
performance as the valves can close with abnormally high velocity
and even bounce off the seat (valve floating and bouncing)
\cite{kura:01,teovorata:06,Choi:00}. To avoid this phenomenon, a
large spring force and preload are applied to the follower
\cite{Raghavacharyulu}. This often causes an increase in the
contact force, which induces higher stresses possibly leading to
early surface failure of the parts. The resulting high friction
valve train reduces the efficiency of the engine system
\cite{Tumer}.

In general, cam-follower systems can be thought of as impact
oscillators with moving boundaries
\cite{Koster:74,Norton:02,Dresner:95,Yan:96}. While the dynamics
of impact oscillators with continuous forcing has been the subject
of many papers in the existing literature (see for example
\cite{peterka2000}, \cite{DaNo:00}, \cite{budd94,budd96}), the
possible intricate bifurcation behaviour of impact oscillators
with discontinuous forcing was discussed only recently, as for
example in \cite{BuPi:06}. It was proposed that discontinuously
forced oscillators can show a novel bifurcation phenomenon unique
to their nature which was termed as corner-impact bifurcation.
Namely, in \cite{BuPi:06} the dynamics are studied of an impact
oscillator forced by a discontinuous sinusoidal forcing of the
form $f(t)=A\vert \sin(\omega t) \vert$. It was shown that, under
variation of the system parameters, abrupt changes of the system
qualitative behaviour are observed when an impact occurs at a
point where the forcing velocity is discontinuous (a corner-impact
bifurcation point).

The observed behaviour was explained in terms of appropriate local
maps. In particular, by using the technique of
discontinuity-mappings recently proposed in \cite{DaNo:00} and
\cite{BeBu:01}, it was suggested that a corner-impact bifurcation
of the oscillator corresponds to a border-collision of a fixed
point of the associated Poincar\'e map. An important difference
was highlighted between corner-impact bifurcations and other types
of discontinuity-induced bifurcations \cite{diBernardo:06} in
impacting systems such as grazing of limit cycles
\cite{No:91,Todd:97,LeCa:00,diBernardo:01b,Piiroinen:04,LeNi:04}.
While the normal form map of a grazing bifurcation is typically
characterised by a square root singularity \cite{No:91}, the local
normal form map associated to a corner-impact bifurcation was
shown to be a piecewise linear map with a gap such as those
studied in \cite{HoHi:06}. Hence, as explained in \cite{BuPi:06},
an appropriate classification method needs to be used to
investigate this novel class of bifurcations.

In \cite{diBernardo:05}, it was conjectured for the first time
that corner-impact bifurcations are fundamental in organizing the
complex behaviour observed in cam-follower systems. It was shown
that, as the cam rotational speed increases, these systems can
exhibit sudden transitions from periodic solutions to chaos. Such
transitions were conjectured to be due to corner-impact
bifurcations.

In this paper, we present a careful analysis of corner-impact
bifurcations in cam-follower systems. We derive analytically the
normal form map associated to such a bifurcation in a
representative example of interest where the cam profile is
characterised by a discontinuous acceleration. In particular, we
investigate the bifurcation behaviour exhibited by this system
under variations of the cam rotational speed. We find that
following the detachment of the follower from the cam, the system
can exhibit complex nonlinear phenomena involving chattering,
period adding cascades and the sudden transition from periodic
attractors to chaos. We explain the sudden transition to chaos
observed in the system in terms of a corner-impact bifurcation.
Namely, we show that dramatic changes in the system behaviour are
observed when, under parameter variation, one of the impacts
characterizing the system trajectory crosses one of the manifolds
in phase space where the cam acceleration is discontinuous.

We prove that the normal form map of the corner-impact bifurcation
in these systems is a piecewise linear continuous map rather than
discontinuous because of the higher degree of discontinuity of the
forcing signal provided by the cam with respect to that of the
forcing considered in \cite{BuPi:06}. We wish to emphasize that
such a finding is generic for the the wide class of impacting
systems characterised by forcing terms with discontinuous
acceleration.

As shown in the paper, the derivation of the mapping has an
immediate practical relevance. In fact, the derivation of a
piecewise linear normal form map implies that the strategy to
classify border-collisions in piecewise linear continuous maps due
to Feigin \cite{BeFe:99} can be used, under some circumstances, to
classify corner-impact bifurcations in continuous-time impacting
flows.

The rest of the paper is outlined as follows. In Sec. 2, we
present the modelling of the cam follower system of our concern
where the cam profile has been assumed to be characterized by a
discontinuous acceleration. Then in Sec. 3 the numerical
bifurcation analysis is presented under variation of the cam
rotational speed. In Sec. 4 we present the analysis of the corner
impact bifurcation phenomenon detected in the system and we
classify the ensuing dynamics by using an appropriately derived
local mapping. Finally, conclusions are drawn in Sec. 5.

\section{Modelling}
The formulation of an appropriate model for a cam-follower system
can be a challenging task for most applications. Various models
with different degrees of complexity have been proposed and
extensively studied. They range from simple models with one
degree-of-freedom (DOF) such as that described in \cite{Koster:74}
to complex models characterised by many DOFs, as for example the
21 DOFs model studied in \cite{Seidlitz} where additional effects
of camshaft torsion and bending, backlash, squeeze of lubricant in
bearings are included. Nevertheless, there is a general agreement
in the literature, confirmed by experiments, that a lumped
parameter single degree-of-freedom model is adequate to represent
the main qualitative features of the dynamic behavior of the
system of interest \cite{Barkan,Koster:74,Akiba,Dresner:95}.

\begin{figure}[h]
\setlength{\unitlength}{1mm}
\begin{picture}(80,80)(-20,0)
  \put (0,0){\mbox{\includegraphics[height=8cm]{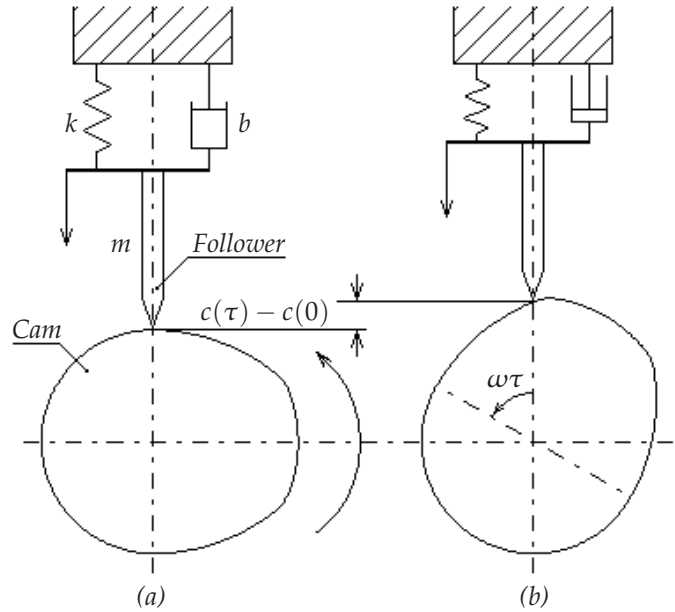}}}
 \put (2,34.5){\mbox{\textit{Cam}}}
 \put (26,46){\mbox{\textit{Follower}}}
 \put (15,46){\mbox{\textit{$m$}}}
 \put (9,62){\mbox{$k$}}
 \put (32,62){\mbox{$b$}}
  \put (27,37){\mbox{$c(\tau)-c(0) $}}
 \put (65,28){\mbox{$\omega \tau$}}
 \put (18.5,-1){\mbox{\textit{(a)}}}
 \put (69.4,-1){\mbox{\textit{(b)}}}
\end{picture}
  \caption{Cam-Follower schematics. \textit{(a)} t=0. \textit{(b)} t=$\tau$.} 
  \label{Fig:cam-shaft}
\end{figure}

The schematic diagram of the cam-follower system under
investigation is shown in Figure \ref{Fig:cam-shaft}. We consider
the following second order equation to model the free body
dynamics of the follower away from the cam,
\begin{eqnarray}
  \label{eq:fb1}
  \nonumber  
    m {q''}(t)+b {q'}(t)+kq(t)=- m g\\
  \text{\textit{if} $q(t)>c(t)$,}
\end{eqnarray}
where $m$, $b$, $k$ and $g$ are constant positive parameters
representing the follower mass, viscous damping, spring stiffness
and the gravitational constant respectively. The state of the
follower is given by the position $q(t)$ and the velocity
${q'}(t)$. The cam position is given by $c(t)$ and we assume that
the follower motion is constrained to the phase-space region where
$q(t) > c(t)$.

The dynamic behavior when impacts occurs (i.e. $q(t) = c(t)$ ) is
modelled via a Newton restitution law as
\cite{Br:99,leine2002,fredriksson1997}:
\begin{eqnarray}
  \label{eq:i1}
   \nonumber  {q'}(t^+) = (1+r){c'}(t)-r {q'}(t^-) \\
  \text{\textit{if} $q(t)=c(t)$,}
\end{eqnarray}
where ${q'}(t^+)$ and ${q'}(t^-)$ are the post- and pre-impact
velocities respectively, ${c'}(t)$ is the projection of the cam
velocity vector at the contact point along the direction of the
free movement of the follower, and $r \in [0,1]$ is the
coefficient of restitution used to model from plastic to elastic
impacts.

\begin{figure}[h]
\setlength{\unitlength}{1mm}
\begin{picture}(65,80)(0,0)
\begin{texdraw}
\drawdim cm \linewd 0.02 \arrowheadtype t:F \arrowheadsize l:0.16
w:0.08
 \move(3.25 4.0) 
 \setgray 0.5 \lpatt() \rmove(-3 0) \ravec(6 0)
 \lpatt() \move(3.25 0.5) \ravec(0 7)
 \linewd 0.03
 \setgray 0
 \move(3.25 4.0) \larc r:2 sd:150.1794 ed:330.1794 
 \move(3.25 4.0) \larc r:2.77 sd: 40.1794 ed:80.1794 
 \move(5.2021    2.8811) \larc r:4.25 sd:117.3427 ed:150.1794 
 \move(1.2979    5.1189) \larc r:4.25 sd:-29.8206 ed:3.0161 
 \move(3.5945    5.9901) \larc r:0.745 sd:80.1794 ed:117.3427 
 \move(4.7931    5.3031) \larc r:0.745 sd: 3.0161 ed:40.1794 
 \linewd 0.02
 \setgray 0.6
 \move(0.6472    5.4919) \avec(5.8528    2.5081) 
 \move(1.5095    0.9634) \avec(4.9905    7.0366) 
 \move(3.25 4.0) \larc r:0.5 sd: 0 ed:60.1794 
 \setgray 0
 \move(3.25 4.0) \fcir f:0 r:0.05 
 \move(5.2021    2.8811)  \fcir f:0 r:0.05 
 \move(1.2979    5.1189)  \fcir f:0 r:0.05 
 \move(3.5945    5.9901) \fcir f:0 r:0.05 
 \move(4.7931    5.3031) \fcir f:0 r:0.05 
 \move(1.5148    4.9946) \fcir f:0.6 r:0.05 
 \move(4.9852    3.0054) \fcir f:0.6 r:0.05 
 \move(3.2500    6.6563)  \fcir f:0.6 r:0.05 
 \move(5.5420    5.3425)  \fcir f:0.6 r:0.05 
 \move(3.7224    6.7291)  \fcir f:0.6 r:0.05 
 \move(5.3661    5.7869)  \fcir f:0.6 r:0.05 
 \textref h:R v:B   \htext(3.200    6.7563){\text{$c(\omega \tau)$}}
 \textref h:C v:C   \htext(4.5 4.3){\text{$\omega \tau = b$}}
 \textref h:C v:C   \htext(3.25 -0.1){\textit{(a)}}
\end{texdraw}
\end{picture}
\begin{picture}(60,80)(-3,0)
 \put (-4,75){\mbox{$c(t)$}}
 \put (-5,50){\mbox{$ {c'}(t)$}}
 \put (-5.5,30){\mbox{$ {c''}(t)$}}
 \put (62,60){\mbox{$t$}}
 \put (30,-1){\mbox{\textit{(b)}}}
 \begin{texdraw}
 \drawdim cm \linewd 0.02 \arrowheadtype t:F \arrowheadsize l:0.16
 w:0.08
 \setgray 1
 \move(0 0) \rlvec(0 1)
 \move(0 0) \rlvec(1 0)
 \setgray 0.2 \lpatt(0.03 0.05)
 \move(1.35 6.35) \rlvec(0 1.2)
 \move(2.0 6.35) \rlvec(0 1.2)
 \move(2.15 6.35) \rlvec(0 1.05)
 \setgray 1
 \move(0 0) \rlvec(0 1)
 \move(0 0) \rlvec(1 0)
 \setgray 0
 \textref h:C v:B   \htext(0.35    6){\textit{a}}
 \textref h:C v:B   \htext(1.15    6){\textit{b}}
 \textref h:C v:B   \htext(1.36    6){\textit{c}}
 \textref h:C v:B   \htext(2.0    6){\textit{d}}
 \textref h:C v:B   \htext(2.2    6){\textit{e}}
 \textref h:C v:B   \htext(3.1    5.85){\textit{f}}
 \textref h:C v:B   \htext(6    6){\textit{a}}
 \end{texdraw}
\end{picture}
  \caption{\text{(a)} Cam profile. \text{(b)} Constraint position $ c(t)$, velocity ${c'}(t)$ and acceleration ${c''}(t)$.}
  \label{Fig:camprofile}
\end{figure}
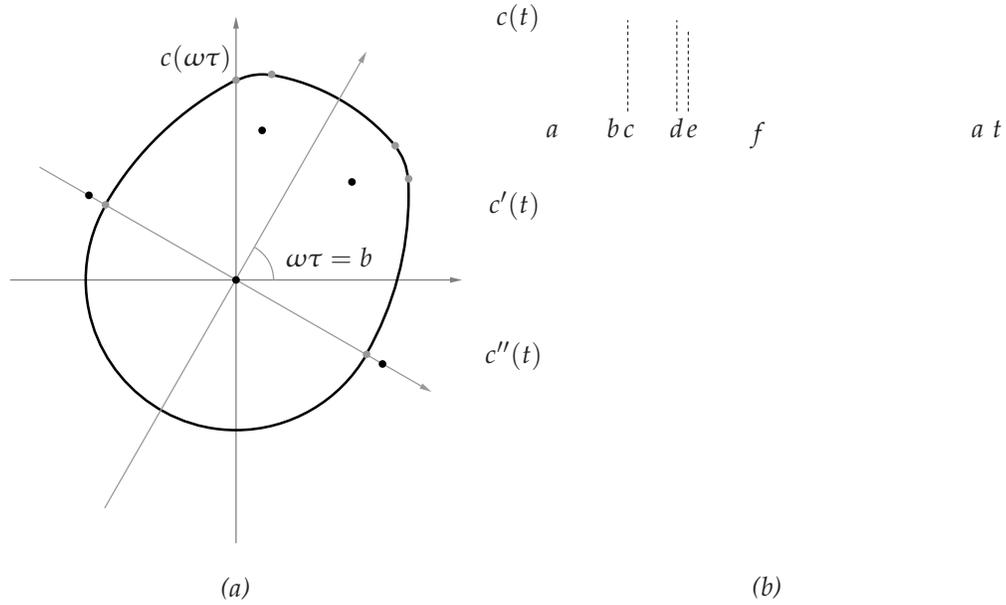

An essential ingredient of the model is the choice of the cam
profile, $c(t)$. The cam is assumed to be rotating at a constant
angular velocity $\omega$ and can be interpreted as the ``control
action'' acting on the follower state as suggested in
\cite{Norton:02}. Therefore, $c(t)$ is carefully selected in
applications as a trade off between several optimality criteria
dependent upon the specific device being considered and the
unavoidable physical constraints present on the system.

Typically, this results in a design process where the cam profile
is selected by using splines and can contain several degrees of
discontinuity. For example,  the cam for a single overhead
camshaft valve train is designed by using quadratic splines and,
as a consequence, discontinuities are present in its acceleration.
In general, it is not uncommon in applications, to find cam
geometries characterised by continuous cam positions and
velocities but a discontinuous second-derivative \cite{Norton:02}.

In what follows, we assume the cam profile to be characterised by
a discontinuous second derivative as shown in Figure
\ref{Fig:camprofile}. For the sake of brevity, the analytical
expressions of the cam profile and its derivatives are reported in
Appendix A. The case of a smooth cam profile with continuous first
and second-order derivatives is also of interest in applications
and was studied experimentally in \cite{AlBe:06}. \label{sec:cam}
\section{Numerical bifurcation analysis}
The model represented by equations \eqref{eq:fb1} and
\eqref{eq:i1} was found to exhibit an intricate bifurcation
behaviour including the sudden transition to chaos under variation
of the cam rotational speed, $\omega$ \cite{diBernardo:05}. The
presence of bifurcations and chaos was also confirmed by
experiments as described in \cite{AlBe:06}.

Here we briefly summarize some of the most striking behaviour
exhibited by the system focusing on the abrupt transition from a
one-periodic impacting solution to chaos observed when $\omega
\approx 673.234445$ rpm.

\begin{figure}[tbhp]
\setlength{\unitlength}{1mm} \centering {
{\begin{picture}(115,45)(0,0)
  \put (0,4.6){\hspace{0.0mm}\mbox{\includegraphics[width=4cm,height=4cm]{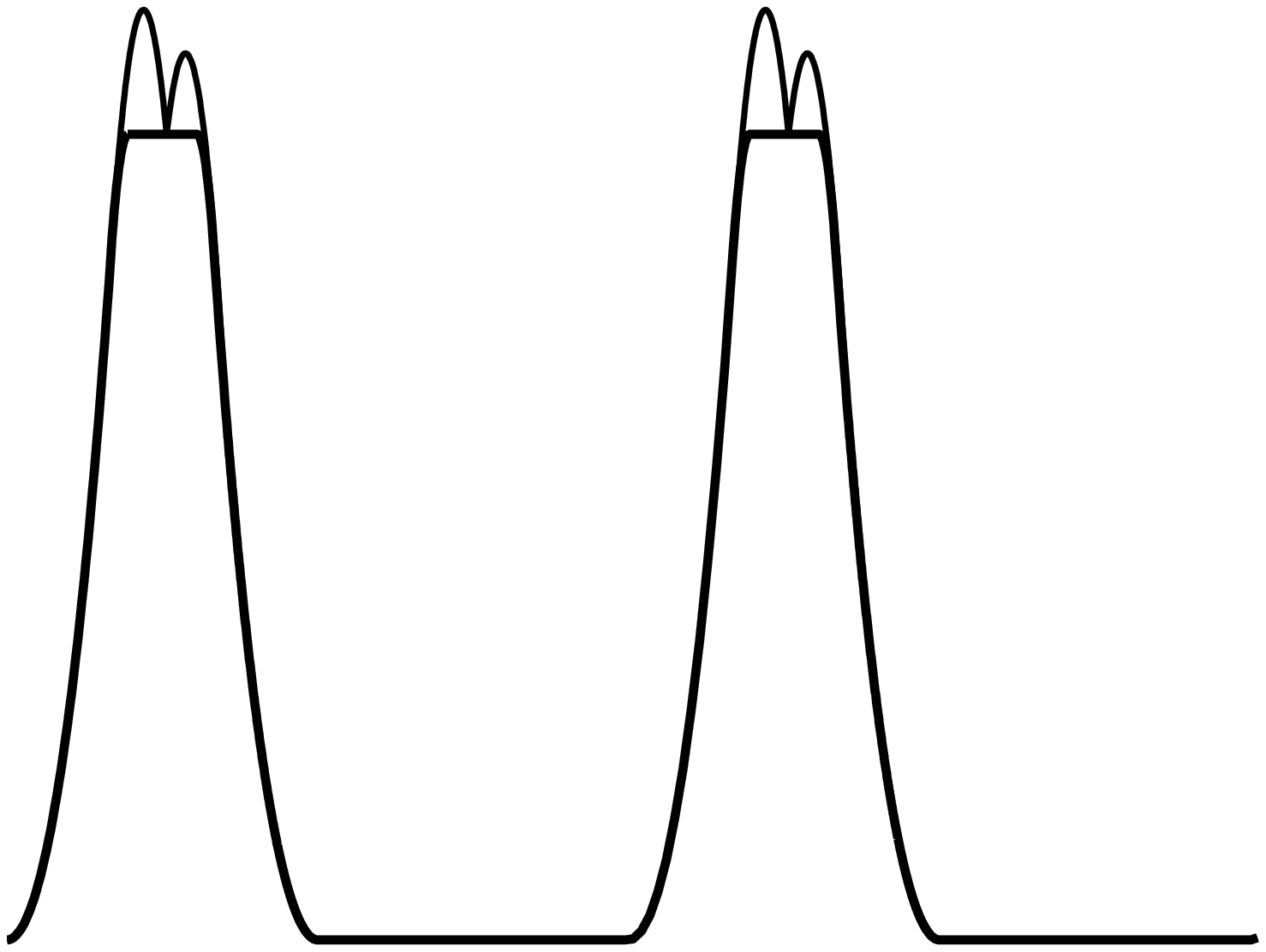}}}
  \put (0,5){\vector(1,0){40}}
  \put (0,5){\vector(0,1){40}}
  \put (38,2){\textit{t}}
  \put (-6,46){\text{$q(t)$}, $c(t)$}
  \put (45,4.9){\hspace{0.0mm}\mbox{\includegraphics[width=4cm,height=3cm]{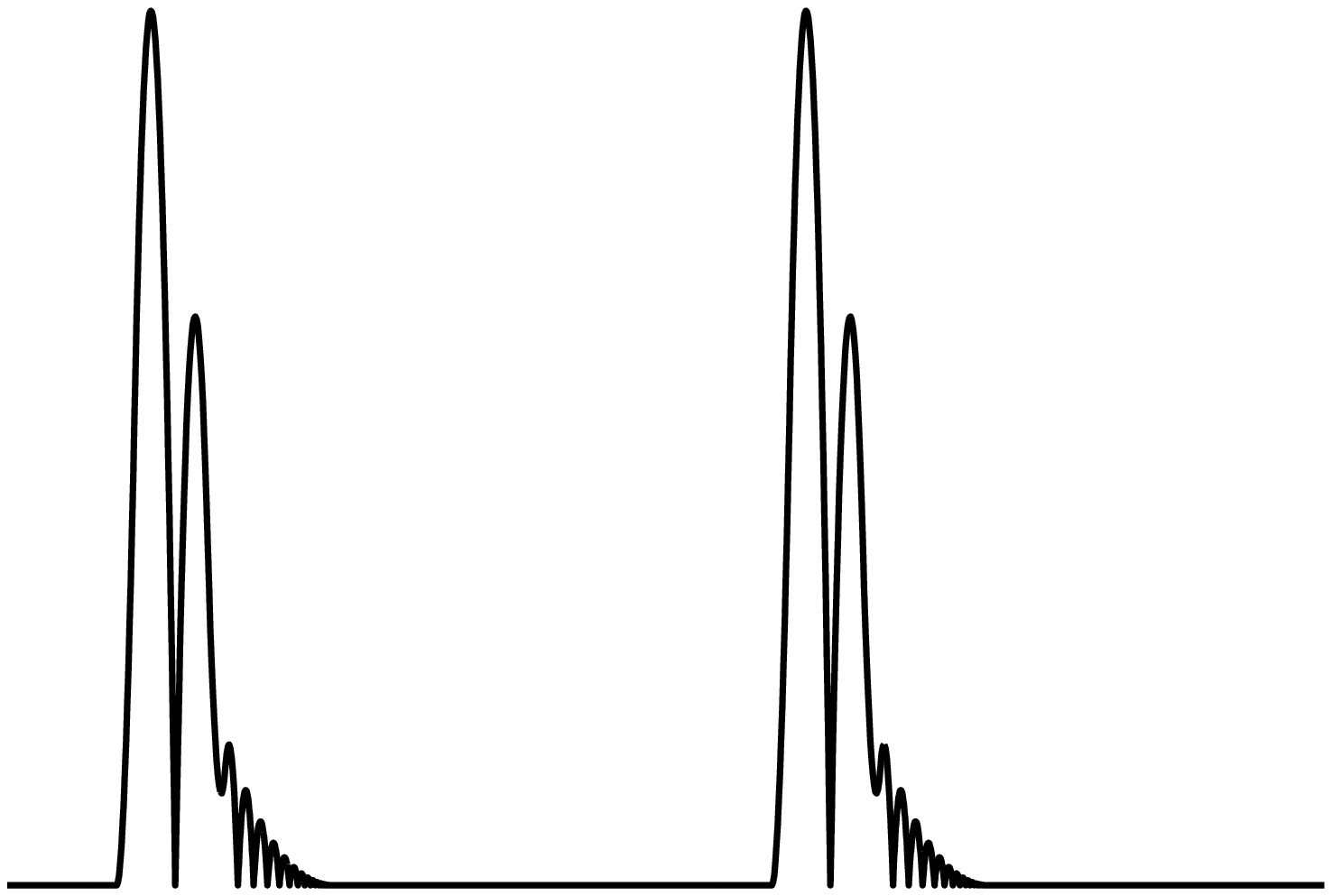}}}
  \put (45,5){\vector(1,0){40}}
  \put (45,5){\vector(0,1){40}}
  \put (83,2){\textit{t}}
  \put (37,46){\text{$q(t)-c(t)$}}
  \put (90,8.5){\hspace{0.0mm}\mbox{\includegraphics[width=2.5cm,height=3cm]{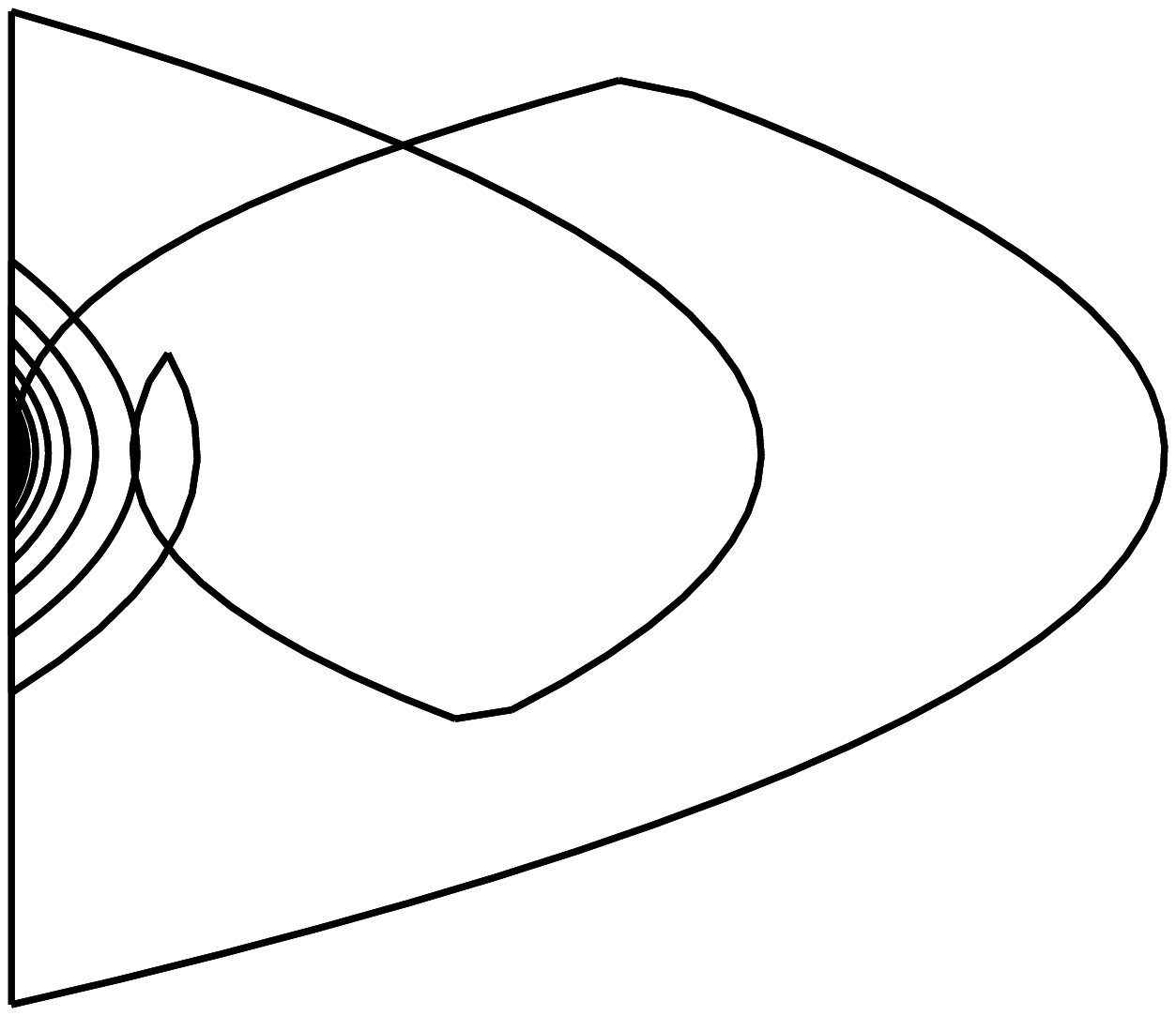}}}
  \put (90,25){\vector(1,0){30}}
  \put (90,5){\vector(0,1){40}}
  \put (90,45){\vector(0,-1){40}}
  \put (115,21){\text{$q(t)-c(t)$}}
  \put (82,46){\text{$ q'(t)- c'(t)$}}
  \put (15,0){{\em(a)}} \put (60,0){{\em(b)}} \put (100,0){{\em(c)}}
\end{picture}}}
  \caption{Time simulation at $\omega = 183$ \textit{rpm}. (a) Follower position, $q(t)$ (Light); Cam position, $c(t)$ (Dark).  (b) Relative position, $q(t)-c(t)$. (c) Phase space, $q(t)-c(t)$ \textit{Vs.} $ q'(t) -  c'(t)$.}
  \label{Fig:timed-cam}
\end{figure}

In general, starting from low values of $\omega$ the system
exhibits solutions characterised by permanent contact between the
cam and the follower. As $\omega$ increases the follower is
observed to detach from the cam during its evolution and then to
impact with it. A typical periodic evolution with impacts is shown
in Fig. \ref{Fig:timed-cam}(a) when $\omega=183$ rpm. We observe
that the follower and the cam are in contact with zero relative
velocity for part of the orbit ({\em sticking}) and then detach
giving rise to impacting behaviour. As shown in
Fig.\ref{Fig:timed-cam}(b)-(c) a careful look at the follower
evolution shows that a {\em chattering sequence} is present, where
theoretically an infinite number of impacts accumulate in finite
time. (Note that in practice chattering is associated to a large
but finite number of impacts.)

\begin{figure}[ht]
\setlength{\unitlength}{1mm}
\begin{picture}(120,170)(0,0)
 \put(4,110){\mbox{\includegraphics*[height=5cm]{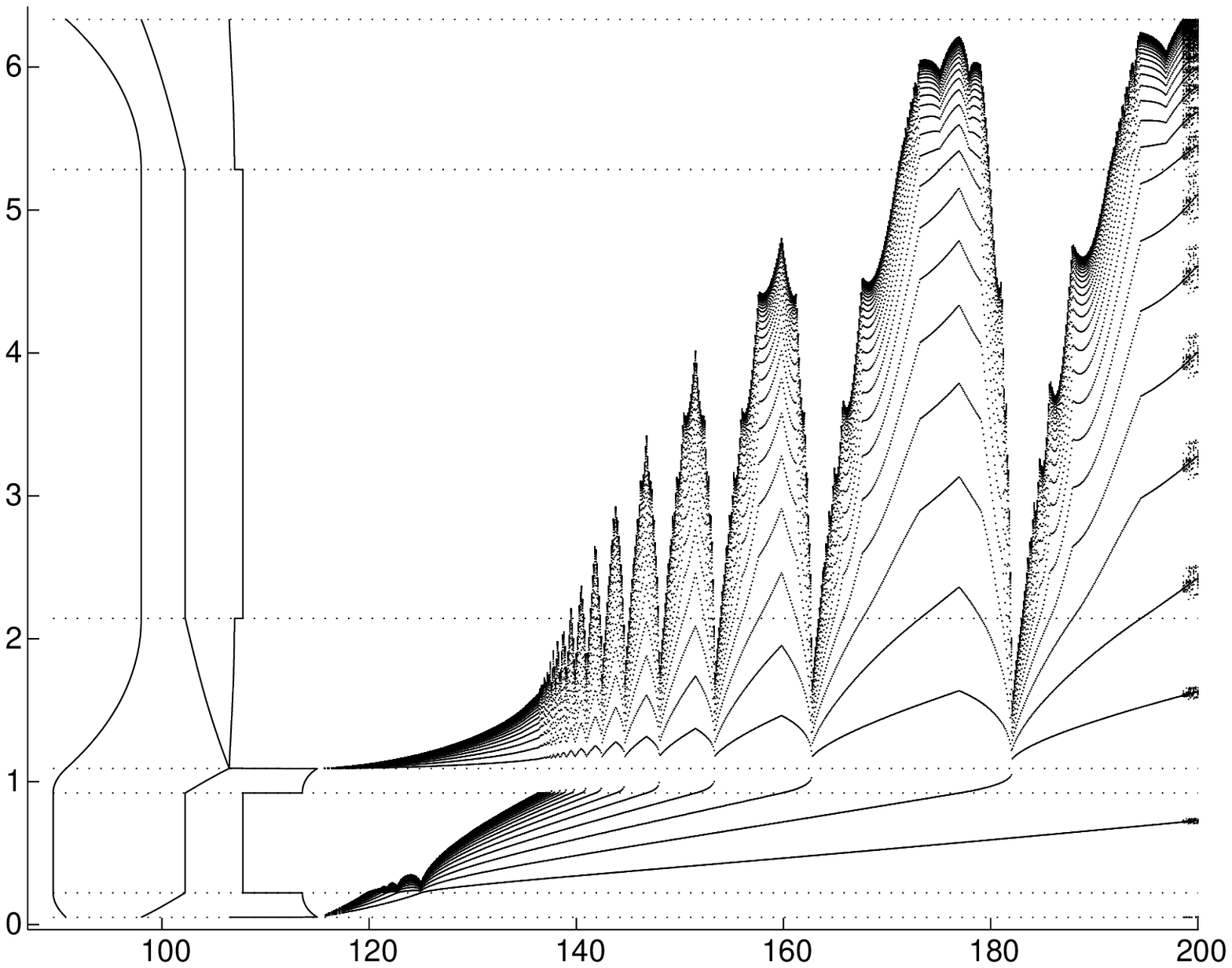}}}
 \put (0,157){{$\phi_i$}} \put (63,107){{$\omega$}}
 \put (68,110){\mbox{\includegraphics*[height=5.5cm]{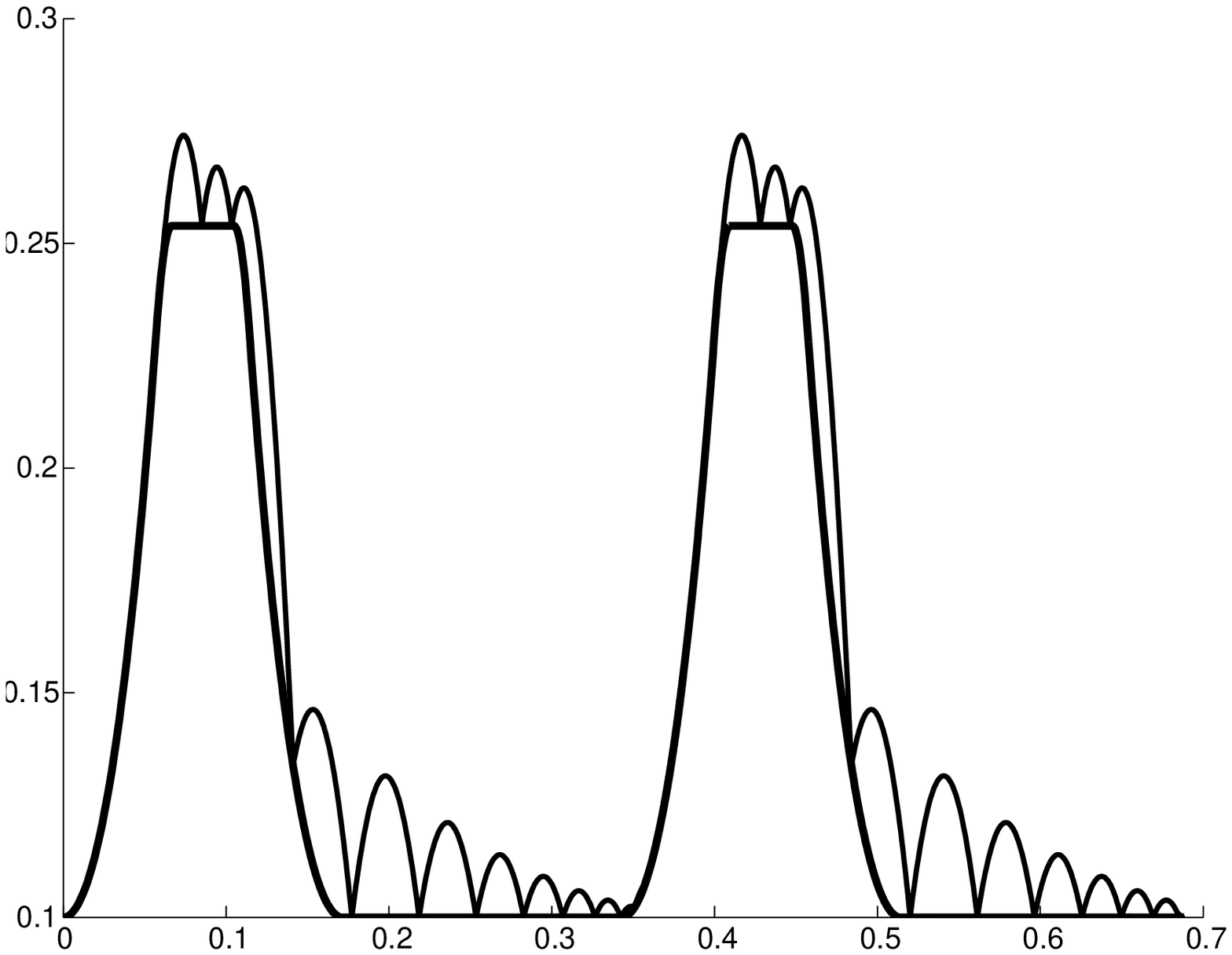}}}
 \put (50.5,110){ \line(0,1){50}}  \put (50.5,116.2){ \circle*{2}}   \put (48,107){{$175$}}
 \put (77.55,149.2){ \circle*{2}}  \put (107.5,149.2){ \circle*{2}}
 \put (130,107){{$t$}}
 \put (3,56){\mbox{\includegraphics*[height=5cm]{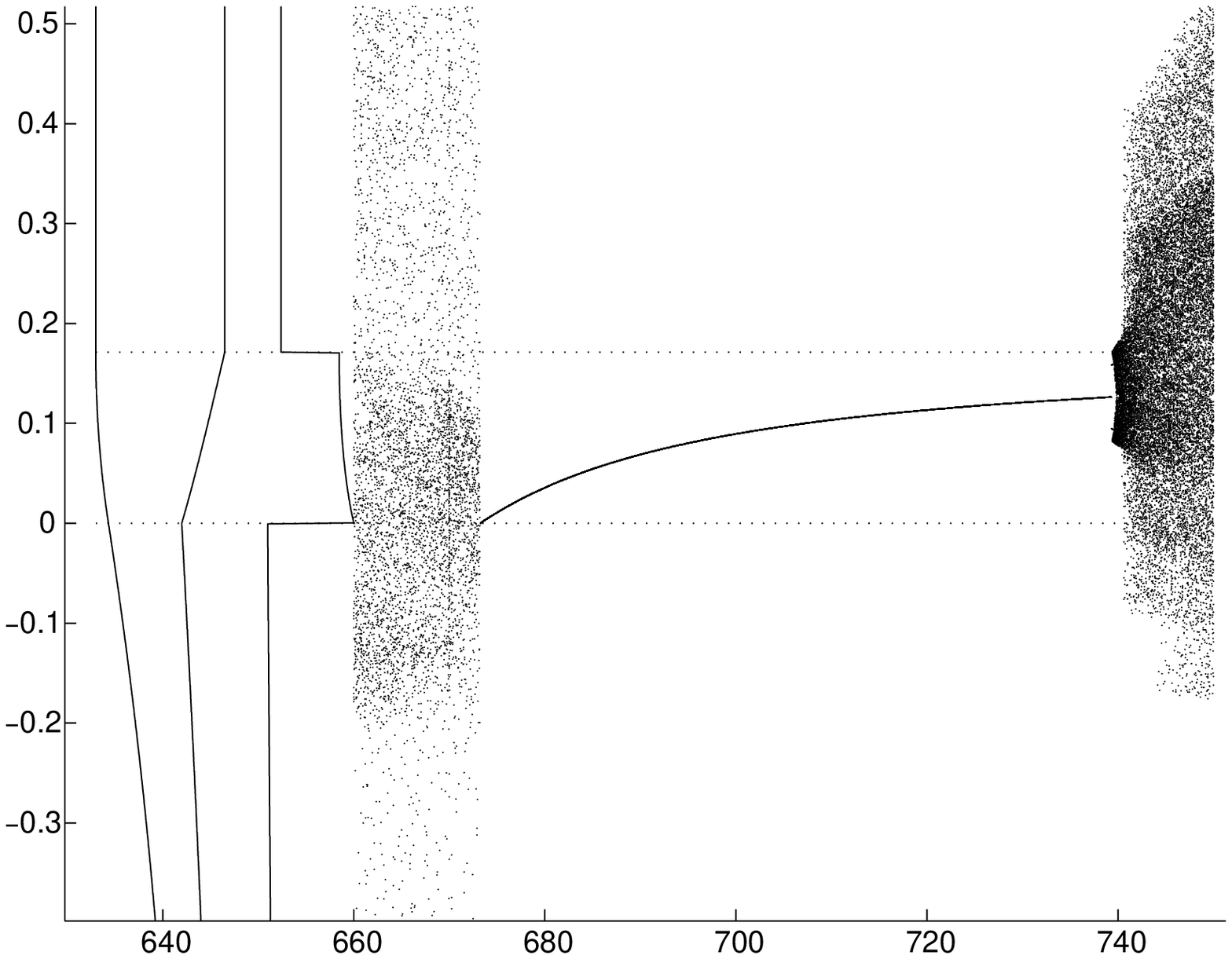}}}
 \put (0,104){{$\phi_i$}} \put (63,54){{$\omega$}}
 \put (68,56){\mbox{\includegraphics*[height=5cm]{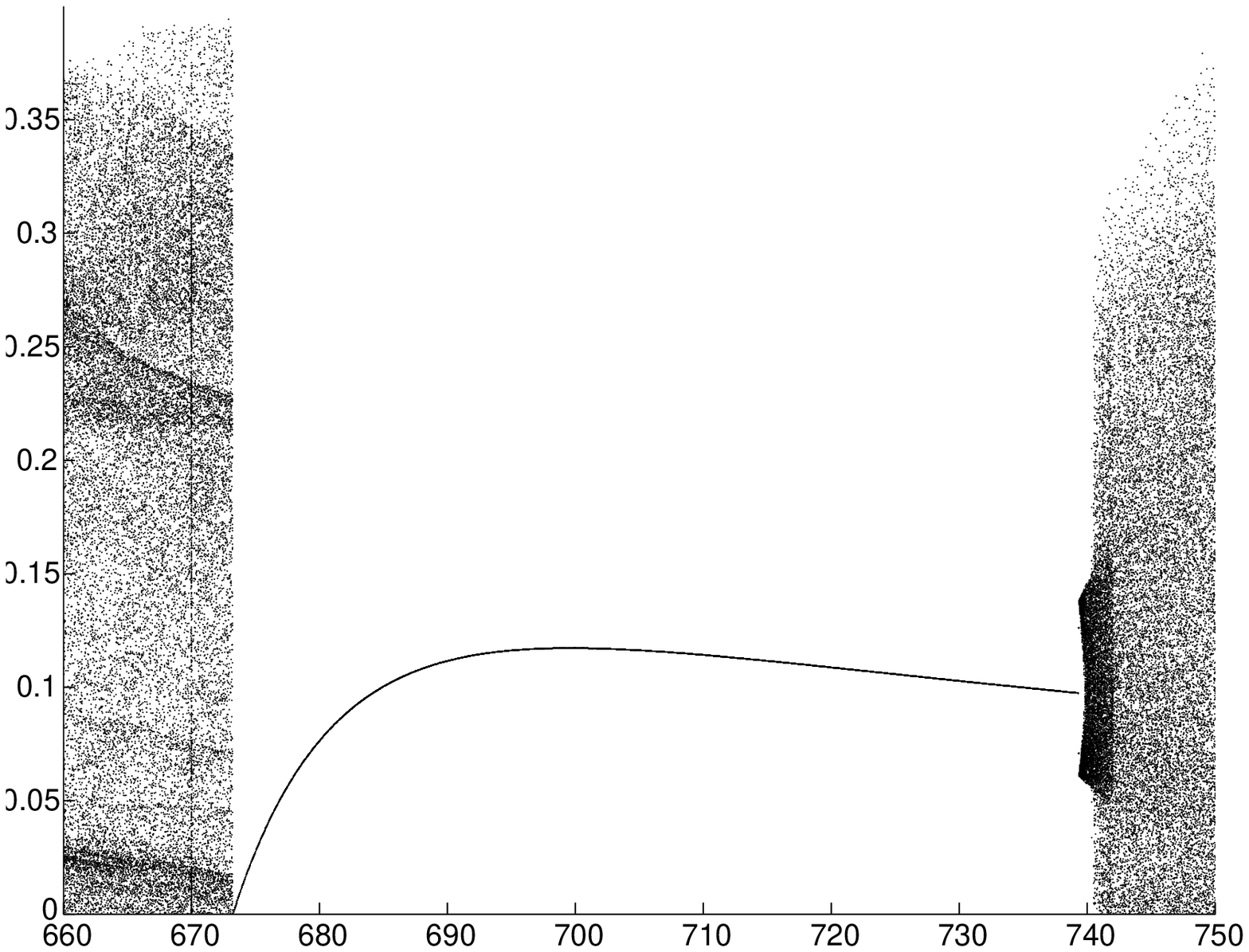}}}
 \put (66.5,105.5){{$q-c$}}\put (130,54){{$\omega$}}
 \put (4,3){\mbox{\includegraphics*[height=5cm]{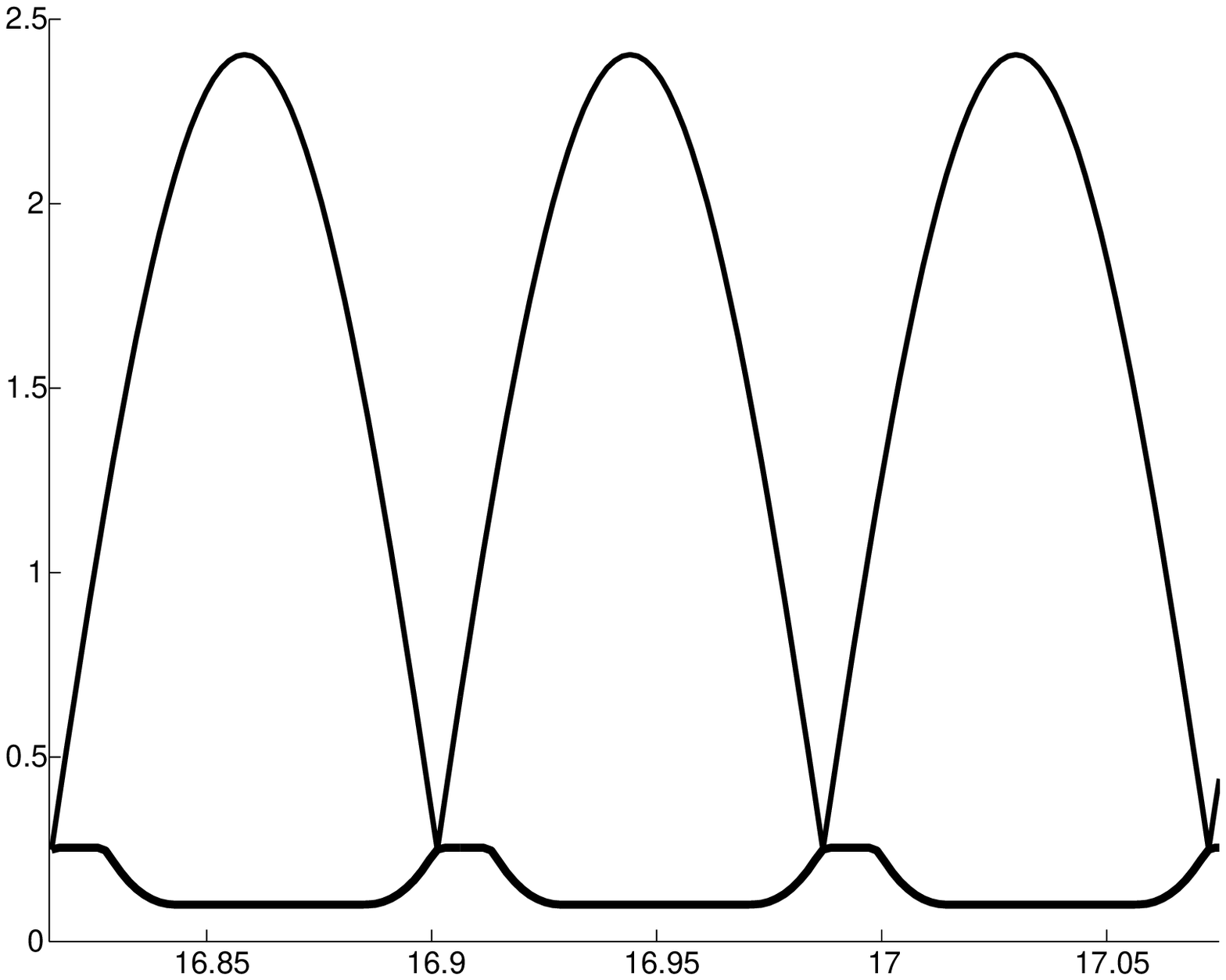}}}
 \put (0,49){{$q,c$}}  \put (63,2){{$t$}}
 \put (69,3){\mbox{\includegraphics*[height=5cm]{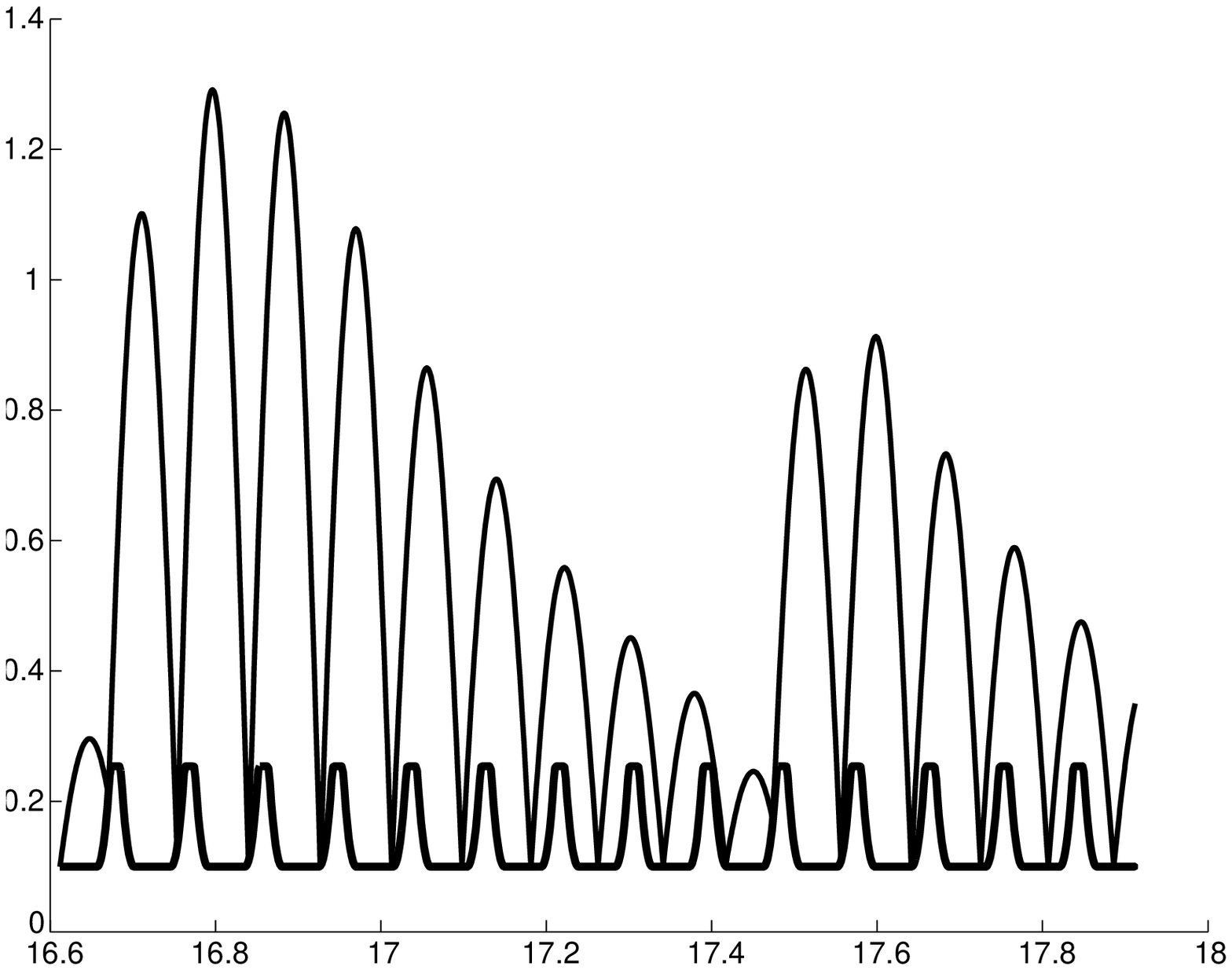}}}
 \put (66,49){{$q,c$}}   \put (130,2){{$t$}}
 \put (35,106){\mbox{\textit{(a)}}}   \put (100,106){\mbox{\textit{(b)}}}
 \put (35,52){\mbox{\textit{(c)}}}    \put (100,52){\mbox{\textit{(d)}}}
 \put (35,0){\mbox{\textit{(e)}}}    \put (100,0){\mbox{\textit{(f)}}}
\end{picture}
\caption{{\em(a)} Impact bifurcation diagram for $[115,200]$
$rpm$. The phase of an impact $\phi_i$ (rad), is plotted agaist
$\omega$. {\em(b)} Time evolution for $175$ $rpm$. {\em(c)} Impact
Bifurcation diagram for $\omega = [660,750]rpm$.{\em(d)}
Stroboscopic Bifurcation diagram for $\omega = [660,750]rpm$.
{\em(e)} Bifurcating orbit at the corner impact point at $\omega =
700 rpm$. {\em(f)} Chaotic evolution for $\omega = 670 rpm$.
Dotted and dashed lines in the diagrams, represent phases where
the cam profile is discontinuous. Vertical curves in panels
(a),(c) shows the cam position velocity and acceleration as
function of the phase.} \label{fig:bif_time}
\end{figure}

Chattering can be associated to an intricate bifurcation
structure. 
In Fig.~\ref{fig:bif_time}(a), the location of the impacts in the
cam surface is depicted for each value of $\omega$, characterising
the follower asymptotic solution. We see that following detachment
at about 114 rpm, the follower immediately exhibits
multi-impacting behaviour and chattering (characterised by the
accumulation of the impact lines in the diagram onto the darker
areas corresponding to the chattering accumulation points). An
interesting phenomenon is the appearance of resonant peaks
associated to impact lines crossing the boundaries where the cam
acceleration profile is discontinuous (represented by dotted lines
in the figure). A detailed analysis of this bifurcation scenario
is presented in \cite{MeOs:06}.

This phenomenon can be classified as due to a {\em corner-impact
bifurcation}, a type of discontinuity-induced bifurcation recently
described in \cite{BuPi:06}. Namely, at certain values of
$\omega$, one of the impacts characterising the follower motion
occurs at a point on the cam profile where the acceleration is
discontinuous. We shall seek to investigate analytically this
phenomenon and classify the behaviour following the corner-impact
event in the cam-follower system of interest. For the sake of
simplicity, we focus on a different region of the system
bifurcation diagram depicted in Fig.~\ref{fig:bif_time}(c). Here a
one-periodic solution characterized by one impact per period
exhibits sudden transitions to chaos as $\omega$ is decreased
below $673.234445$ rpm. A close look at the impact bifurcation
diagram in Fig.~\ref{fig:bif_time}(c) and in the stroboscopic
bifurcation diagram Fig.~\ref{fig:bif_time}(d) shows that such
transitions occur precisely when the impact characterising the
solution crosses the cam discontinuity boundaries (the dotted
lines in figure ~\ref{fig:bif_time}(c)). Specifically, the sudden
transition to chaos is due to the corner-impact bifurcation of the
periodic solution depicted in Fig.~\ref{fig:bif_time}(e). Past the
corner-impact bifurcation point the system exhibits chaotic
behaviour (see for example the trajectory reported in
Fig.~\ref{fig:bif_time}(f) for $\omega \approx 670$ rpm). The rest
of this paper is devoted to the analysis of this bifurcation
scenario.
\clearpage

\section{Corner Impact Bifurcation Analysis}
The numerical observations reported above indicate that a
corner-impact bifurcation is causing the transition to chaos
observed in the cam-follower system. Specifically, we are
interested in analyzing the occurrence of the corner impact
bifurcation depicted in Fig.~\ref{fig:bif_time}(c) when \linebreak
$\omega \approx 673.234445$ rpm. Numerically, we detected that the
bifurcating orbit, shown in Fig.~\ref{fig:bif_time}(e) is a
one-periodic orbit characterised by one impact per period. As the
rotational speed of the cam is decreased, at the bifurcation
point, the impact is observed to cross the point on the cam
surface where the cam acceleration is discontinuous. To
investigate this novel type of discontinuity-induced bifurcation
we will construct analytically the Poincar\'e map of the system
close to the bifurcation point. We will then study the
bifurcations of the fixed point corresponding to the periodic
solution of interest. A crucial point in the analysis is to assess
whether the resulting map is piecewise linear continuous or not.
Indeed, only if this is the case, the theory of border-collision
bifurcations (see \cite{BaGr:99, BeFe:99}) can be used to classify
the possible solutions branching from the corner-impact
bifurcation point \cite{diBernardo:01}.

We use the concept of discontinuity mapping (or normal form map)
recently introduced in \cite{DaNo:00}, \cite{BeBu:01} to construct
analytically the Poincar\'e map associated to the bifurcating
orbit of interest. We use the cam-follower system described in Sec
\ref{sec:cam} as a representative example to carry out the
analytical derivations.

\subsection{Poincar\'e Map Derivation}
\begin{figure}[hbtp]
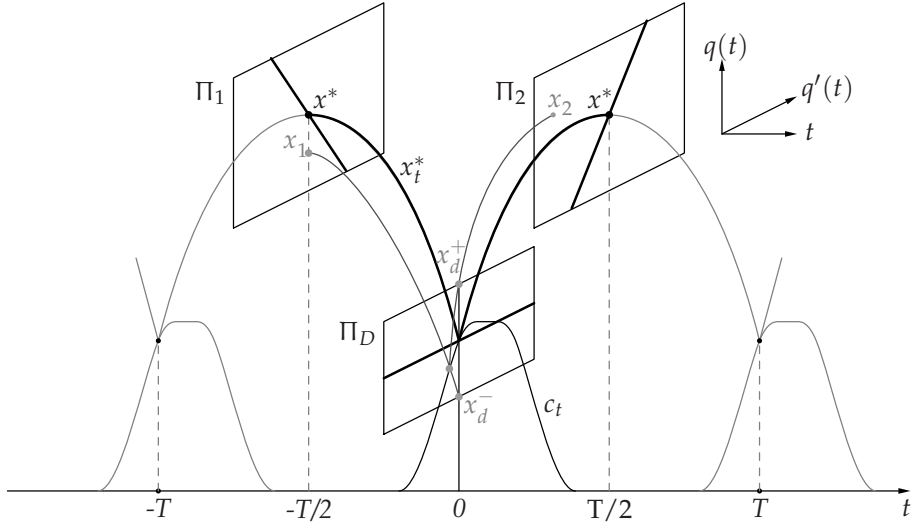

\begin{center}
\begin{texdraw}
\drawdim cm \linewd 0.02 \arrowheadtype t:F \arrowheadsize l:0.16
w:0.08
\move(3.5 4.75) \ravec(1 .5) \rmove(-1 -0.5) \ravec(0 1)
 \rmove(0 -1) \ravec(1 0)
\move(0 0) \rlvec(0 2.75) \move(-6 0) \avec(6 0)
 \move(-3 3.5) \rlvec(2 1) \rlvec(0 2) \rlvec(-2 -1) \rlvec(0 -2)
 \move(1 3.5) \rlvec(2 1) \rlvec(0 2) \rlvec(-2 -1) \rlvec(0 -2)
 \move(-1 0.75) \rlvec(2 1) \rlvec(0 1.5) \rlvec(-2 -1) \rlvec(0 -1.5)
 \lpatt(0.1 0.1) \setgray 0.5 \linewd 0.02
 \move(-2 0) \rlvec(0 5) \lpatt() \setgray 0 \linewd 0.03 \rmove(-0.5 0.75) \rlvec(1 -1.5)
 \lpatt(0.1 0.1) \setgray 0.5 \linewd 0.02
 \move(2 0) \rlvec(0 5) \lpatt() \setgray 0 \linewd 0.03 \rmove(0.5 1.25) \rlvec(-1 -2.5)
 \setgray 0 \linewd 0.03 \move(-1 1.5) \rlvec(2 1)
 \lpatt() \setgray 0.5 \linewd 0.02
 \move(-4.8 0)  \clvec(-4.5 0)(-4.2 1.5)(-4 2) \clvec(-3.9
 2.25)(-3.8 2.25)(-3.75 2.25) \rlvec(0.25 0) \clvec(-3.45
 2.25)(-3.35 2.25)(-3.25 2) \clvec (-3.05 1.5)(-2.75 0)(-2.45 0)
 \setgray 0 \lpatt() \move(-0.8 0)  \clvec(-0.5 0)(-0.2 1.5)(-0 2)
 \clvec(0.1 2.25)(0.2 2.25)(0.25 2.25) \rlvec(0.25 0) \clvec(0.55
 2.25)(0.65 2.25)(0.75 2) \clvec (0.95 1.5)(1.25 0)(1.55 0)
 \setgray 0.5 \lpatt() \move(3.2 0)  \clvec(3.5 0)(3.8 1.5)(4 2)
 \clvec(4.1 2.25)(4.2 2.25)(4.25 2.25) \rlvec(0.25 0) \clvec(4.55
 2.25)(4.65 2.25)(4.75 2) \clvec (4.95 1.5)(5.25 0)(5.55 0)
 \lpatt() \setgray 0.5 \move(-4 2) \clvec(-3 6)(-1 6)(0 2)
 \lpatt() \setgray 0.5 \move(0 2) \clvec(1 6)(3 6)(4 2) \lpatt()
 \setgray 0 \linewd 0.03 \move(-2 5) \clvec(-1.2 5)(-0.5 4)(0 2 )
 \clvec(0.5 4)(1.2 5)(2 5) \linewd 0.02
 \lpatt() \setgray 0.3 \move(-2 4.5) \clvec(-1.5 4.5)(-0.5 3)(0
 1.25) \move(-2 4.5) \fcir f:0.6 r:0.05 \move(0 1.25) \fcir f:0.6
 r:0.05
 \lpatt() \setgray 0.3 \move(-0.125 1.625) \clvec(0 4)(0.75
 4.75)(1.25 5) \move(-0.125 1.625) \fcir f:0.6 r:0.05 \move(1.25 5)
 \fcir f:0.6 r:0.03 \move(0 2.75) \fcir f:0.6 r:0.05
 \lpatt() \setgray 0.5 \move(-4 2) \rlvec(-0.3 1.1) \move(4 2)
 \rlvec(0.3 1.1)
\move(4 0) \fcir f:0 r:0.03 \lpatt(0.1 0.1) \rlvec(0 2) \fcir f:0
 r:0.03
 \move(-4 0) \fcir f:0 r:0.03 \lpatt(0.1 0.1) \rlvec(0 2) \fcir f:0
 r:0.03
 \textref h:C v:T   \htext(-4 -0.1){\textit{-T}}
                    \htext(-2 -0.1){\textit{-T/2}}
                    \htext(0 -0.1){\textit{0}}
                    \htext(2 -0.1){T/2}
                    \htext(4 -0.1){\textit{T}}
 \textref h:R v:T   \htext(6 -0.1){\textit{t}}
 \textref h:R v:C   \htext( -3.1 5.3){$\Pi_{1}$}
                    \htext( -1.1 2.1){$\Pi_{D}$}
                    \htext(0.9 5.3){$\Pi_{2}$}
 \textref h:R v:B   \htext( -2 4.45){\color[gray]{0.5}${x}^{}_1$}
 \textref h:C v:B   \htext( 1.35 5){\color[gray]{0.5}${x}^{}_2$}
 \textref h:L v:T   \htext( 0.05 1.35){\color[gray]{0.5}${x}^{-}_d$}
 \textref h:C v:B   \htext( -0.1 2.9){\color[gray]{0.5}${x}^{+}_d$}
 \textref h:L v:B   \htext( -0.8 4.1){${x}^{*}_t$}
 \textref h:C v:B   \htext( 1.25 1){$c_t$}
 \textref h:L v:B   \htext( -1.95 5.1){$x^*$}
 \move(-2 5) \fcir f:0 r:0.05
 \textref h:R v:B   \htext( 2.05 5.1){$x^*$}
 \move(2 5) \fcir f:0 r:0.05
 \textref h:C v:T   \htext( 3.55 6.1){$q(t)$}
 \textref h:L v:T   \htext( 4.55 5.55){$q'(t)$}
 \textref h:L v:T   \htext( 4.6 4.9){$t$}
\end{texdraw}
\end{center}
 \caption{Global map composition.}
  \label{Fig:globalmap_2b}
\end{figure}

We are interested in the analysis of the period one orbit at the
corner-impact bifurcation point. Such orbit is sketched in figure
\ref{Fig:globalmap_2b}. Then, close to such periodic orbit we
define the stroboscopic map $P$ as the mapping from the follower
state $x_1 \in \Pi_1$ at a stroboscopic time instant $t_1$ to the
next stroboscopic point $x_{2} \in \Pi_2$. Without loss of
generality, we assume that $t_n=-\frac{T}{2}+(n-1)T$ for $n=1,2,3,
\ldots$, where $T$ is the period of the cam forcing cycle (note
that $T=2\pi / \omega$). Namely, we have:
\begin{eqnarray}
\label{eq:lmap_g} x_{2}=P(x_{1}).
\end{eqnarray}
To construct $P$ we would need to flow forward using the system
evolution from $x_1$ to $x_{2}$ for time $T$ taking into account
the possible occurrence of impacts and therefore applying Newton's
restitution law as required. Alternatively, as shown in
\cite{DaNo:00}, it is possible to construct $P$ as the composition
of three submappings: (i) an affine transformation $P_{1,T/2}$
from the stroboscopic plane $\Pi_1$ at $t_1=-\frac{T}{2}$ to the
plane $\Pi_D$ going through the corner impact point at $t=0$; (ii)
an appropriate  zero-time discontinuity mapping(ZDM) $P_D$ on
$\Pi_D$ accounting for the presence of the discontinuity; and
again (iii) an affine transformation $P_{2,T/2}$ from the plane
$\Pi_D$ at $t=0$ back to the stroboscopic plane $\Pi_2$ at
$t_2=\frac{T}{2}$. Specifically, while $P_{1,T/2}$ and $P_{2,T/2}$
are fixed time maps that accounts for the follower evolution away
from the cam as if no impact had occurred, the ZDM represents the
correction that needs to be made to the system trajectories
because of the presence of impacts. Figure \ref{Fig:globalmap_2b}
represents the global map composition. This means that we can
write
\begin{equation}P = P_{2,T/2}\circ P_D \circ
P_{1,T/2},\end{equation} where $P_{1,T/2} : \Pi_1 \mapsto
\Pi_{D}$, will map the state from the initial condition $x_1$ on
the stroboscopic plane $\Pi_1$ to a point ${x}^{-}_{d}$ on the
discontinuity plane $\Pi_{D}$ as if no impacts had occurred. $P_D:
\Pi_D \mapsto \Pi_{D}$ will then map ${x}^{-}_{d}$ to the point
${x}^{+}_{d}$ appropriately correcting the evolution for the
presence of impacts (See Fig. \ref{Fig:ZDM5}). Finally $P_{2,T/2}
: \Pi_D \mapsto \Pi_{2}$, \linebreak will map ${x}^{+}_{d}$ to a
point ${x}_{2}$ back onto the stroboscopic plane $\Pi_{2}$. In so
doing, as discussed in \cite{DaNo:00}, \cite{BeBu:01}, the effect
of the system discontinuities due to impacts are all taken into
account by the ZDM, $P_D$, which is therefore often termed as the
local normal form map in the context of the theory of
discontinuity-induced bifurcations \cite{diBernardo:01b}.

\subsubsection{Derivation of $P_{1,T/2}$ and $P_{2,T/2}$}
As explained above, the maps $P_{1,T/2}$ and $P_{2,T/2}$ are
defined only in terms of the free body dynamics of the follower
and the cam rotating period $T$ (depending upon the cam rotational
speed $\omega$). Therefore we can solve equations \eqref{eq:fb1}
to get an analytical expression of the flows generating the
mappings of interest.

Specifically, we define
$${x}_t=\left[\begin{array}{c}q(t)+\frac{g}{{\omega}_{0}^2}\\
q'(t)\end{array}\right],\hspace{5mm}y_t=\left[\begin{array}{c}c(t)\\
 c'(t)\end{array}\right].$$
 as the state vector for the follower
and the cam respectively.

Then, as explained in Appendix B, the generalized solution of
\eqref{eq:fb1} is:
\begin{eqnarray}
\label{eq:fbsol1}{x_t}&=&e^{-\zeta t} \left({I}\cos(\omega_s
t)+{A}\sin(\omega_s t)\right){x_0} \label{eq:gff}\\ \nonumber
&=&\phi_t x_0,
\end{eqnarray}
where  $\zeta=\frac{b}{2m}$, $\omega_0=\sqrt{\frac{k}{m}}$,
$\omega_s={\sqrt{\omega_0^2-\zeta^2}}$, $I$ is the identity
matrix, $\phi_t x_0$ represents the system flow for time $t$
starting from the initial condition $x_0$ and
$${A}=\left[\begin{array}{cc}\frac{\zeta}{\omega_s}&\frac{1}{\omega_s}\\
-\frac{{\omega}_{0}^2}{\omega_s}&-\frac{\zeta}{\omega_s}\end{array}\right].$$
Note that, in general, the system flow operator can be expressed
as:
\begin{eqnarray}
\phi_t=\frac{e^{-\zeta
t}}{\omega_s}\left[\begin{array}{cc}\omega_s\cos(\omega_s t
)+{\zeta}\sin(\omega_s t )&\sin(\omega_s t
)\\-{{\omega}_{0}^2}\sin(\omega_s t )&\omega_s\cos(\omega_s t
)-{\zeta}\sin(\omega_s t )\end{array}\right].
\end{eqnarray}
The submapping $P_{i,T/2}$ can then be easily obtained using
\eqref{eq:gff} as:
\begin{eqnarray}
\nonumber P_{i,T/2}(x)&=&e^{-\zeta T/2} \left({I}\cos(\omega_s
T/2)+{A}\sin(\omega_s T/2)\right)x \\
&:=&\phi_{\frac{T}{2}} x. \label{eq:lmaptheta_1}
\end{eqnarray}

\subsubsection{Derivation of $P_D$}
\begin{figure}[hbtp]
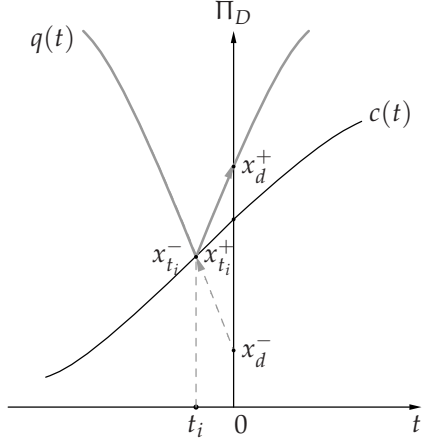

\begin{center}
\begin{texdraw}
\drawdim cm \linewd 0.02 \arrowheadtype t:F \arrowheadsize l:0.16
w:0.08
\move(0 0) \ravec(0 5) \move(-3 0) \avec(2.5 0) \lpatt()
 \lpatt()
 \move(-2.5 0.4)  \clvec(-2.2 0.5)(-1.8 0.7)(-0 2.5) \clvec(1.2 3.6)(1.5 3.7)(1.7 3.8)
 \linewd 0.03 \lpatt() \setgray 0.6 \move(-2 5) \clvec(-1.8
 4.8)(-1.5 4.5)(-0.5 2)
 \linewd 0.02 \lpatt(0.1 0.1) \setgray 0.6
 \move(-2 5) \clvec(-1.8 4.8)(-1.4 4.4)(0 0.75)
 \arrowheadsize l:0.2 w:0.1 \lpatt(0 0.2) \move(0 0.75) \ravec(-0.5 1.25)
 \ravec(0.5 1.24)
 \arrowheadsize l:0.26 w:0.08
 \linewd 0.03 \lpatt() \setgray 0.6 \move(-0.5 2) \clvec(0.5
 4.5)(0.8 4.8)(1 5) \linewd 0.02
\move(-0.5 0) \fcir f:0 r:0.03 \lpatt(0.1 0.1) \rlvec(0 2) \fcir
f:0 r:0.025
 \move(0 0.75) \fcir f:0 r:0.025
 \move(0 3.2) \fcir f:0 r:0.025
 \move(0 2.5) \fcir f:0 r:0.025
 \textref h:C v:T \htext(0.1 -0.1){{$0$}}
 \textref h:C v:T \htext(-0.5 -0.1){$t_i$}
 \textref h:R v:T \htext(2.5 -0.1){{$t$}}
 \textref h:L v:C \htext(0.1 0.75){${x}^{-}_d$}
 \textref h:R v:C \htext(-0.65 2.0){${{x}^{-}_{t_i}}$}
 \textref h:L v:C \htext(-0.4 2.0){${{x}^{+}_{t_i}}$}
 \textref h:L v:C \htext(0.1 3.2){$ {x}^{+}_{d}$}
 \textref h:L v:C \htext(1.8 3.9){$c(t)$}
 \textref h:L v:C \htext(-2.7 4.9){$q(t)$}
 \textref h:C v:C \htext(0 5.25){$\Pi_D$}
\end{texdraw}
\end{center}
 \caption{ZDM construction.}
 \label{Fig:ZDM5}
\end{figure}

As explained in \cite{BeBu:01}, the ZDM can be obtained by an
appropriate composition of backward and forward flows so that the
overall time spent following backward and forward is zero. As
explained earlier, the ZDM is the correction that maps the point
${x}^{-}_{d} \in \Pi_{D}$ onto the point ${x}^{+}_{d} \in \Pi_{D}$
taking into account the presence of impacts in the trajectory of
interest. In what follows we assume that only one impact occurs
over one cycle of the periodic orbit of interest as we suppose to
be sufficiently close to the bifurcating orbit ${x}^{*}_t$ shown
in Fig. \ref{Fig:globalmap_2b}. Figure \ref{Fig:ZDM5} shows a
schematic diagram that describes the construction of the ZDM,
close to the corner-impact bifurcations. Without loss of
generality we assume that the origin is placed at the Poincar\'e
section $\Pi_D$. To derive analytically the mapping
${x}^{+}_{d}=P_D({x}^{-}_{d})$ we need to perform the following
steps:

\begin{enumerate}
\item Starting from ${x}^{-}_d$, we find the time $t_i$ at which the impact
occurs. Namely, $t_i$ is obtained by looking at the difference,
$(q(t)-c(t))$, between the follower position and the cam position
close to $t=0$. Given a vector $z$, we indicate by
$\left[z\right]_1$ its first component. Then
$q(t)=\left[x_t\right]_1 -\frac{g}{{\omega}_{0}^2}$ and therefore,
close to ${x}^{-}_d$, $t_i$ can be obtained as the nearest
solution of the equation:
\begin{eqnarray}
H({x}^{-}_{-t_i}, t_i) := \left[{x}^{-}_{-t_i}-
y_{-t_i}\right]_1=h \cdot \left[\phi_{-t_i} {x}^{-}_d -
y_{-t_i}\right]= 0,
\label{eq:H}
\end{eqnarray}
where $h=[\begin{array}{cc}1 &0\end{array}]$.

Hence, $t_i$ is implicitly defined by the equation
$H({x}^{-}_{-t_i},t_i)=0$. Once, $t_i$ is found, the pre-impact
state of the system, ${x}^{-}_{-t_i}$, can also be obtained as
\begin{eqnarray}
 {x}^{-}_{-t_i} &=& \phi_{-t_i}{x}^{-}_{d}.
 \label{eq:impactdynamics_1}
\end{eqnarray}
Note that $t_i$ can be either negative or positive according to
whether the impact occurs to the left or to the right of $t=0$.
\item  Using the restitution law \eqref{eq:i1}, we can then write the post-impact
state of the follower ${x}^{+}_{-t_i}$ as
\begin{eqnarray}
{x}^{+}_{-t_i}={x}^{-}_{-t_i}+R({x}^{-}_{-t_i}-y_{-t_i})=\rho({x}^{-}_{-t_i},y_{-t_i}),
\label{eq:impactdynamics_2}
\end{eqnarray}
where
$$R=\left[\begin{array}{cc}0&0\\0&-(1+r)\end{array}\right].$$
\item Finally, to obtain ${x}^{+}_{d}$, we flow forward for time $t_i$  starting from the post-impact state ${x}^{+}_{-t_i}$ found at the previous
step. In so doing, the state of the follower $x^{+}_{d} \in
\Pi_{D}$ can be computed as:
\begin{eqnarray} {x}^{+}_{d}=\phi_{t_i}{x}^{+}_{-t_i}. \label{eq:impactdynamics_3}\end{eqnarray}
\end{enumerate}

Using equations
\eqref{eq:impactdynamics_1},\eqref{eq:impactdynamics_2} and
\eqref{eq:impactdynamics_3} we can then write explicitly the ZDM
as:
\begin{eqnarray}
 \label{eq:lmapd}
 {x}^{+}_{d}=& P_D({x}^{-}_{d})&=\left(\mathbf{I}+\phi_{t_i}R\phi_{-t_i}\right){x}^{-}_{d}-\phi_{t_i}R
 y_{-t_i},
\end{eqnarray}
with $t_i$ defined implicitly by the equation \eqref{eq:H}.

\subsubsection{Constructing the Stroboscopic Map}
Composing the submappings $P_{1,T/2}$, $P_{2,T/2}$ and $P_D$ given
by \eqref{eq:lmaptheta_1}
and \eqref{eq:lmapd}, we can then construct the stroboscopic
Poincar\'e map, $P$, of the system close to the corner-impact
bifurcation point from a generic $x_n \in \Pi_n$ to $x_{n+1} \in
\Pi_{n+1}$ as:
\begin{eqnarray}
\nonumber x_{n+1}=P(x_n,T)&=&
P_{2,T/2}(P_D(P_{1,T/2}(x_n)))\\
&=&\phi_{\frac{T}{2}}\left(\left(\mathbf{I}+\phi_{t_i}R\phi_{-t_i}\right)\phi_{\frac{T}{2}}{x}_n-\phi_{t_i}R
 y_{-t_i}\right), \label{eq:lmap}
\end{eqnarray}
where $t_i$ is implicitly defined by the equation $H(x_n,t_i)=h
\cdot \left(\phi_{\frac{T}{2}-t_i} x_n - y_{-t_i}\right) =0$.

Note that the fixed point ($x^*$ associated to the periodic
solution existing for a fixed value of the cam period $T=T^*$),
can be obtained by solving equation \eqref{eq:lmap} for
$x_{n+1}=x_{n}={x}^{*}$ i.e.,
\begin{eqnarray}
 {x}^{*}=
 -\left[\mathbf{I}-\phi_{T^*}+\phi_{\frac{T^*}{2}}R\phi_{\frac{T^*}{2}}\right]^{-1}\phi_{\frac{T}{2}}Ry_{0},
 \label{eq:fixedp}
\end{eqnarray}
with ${t}^*_i=0$.

In what follows we are interested in studying such mapping locally
to the corner-impact bifurcation point detected when
$\omega=\omega^*=673.234445$ rpm, corresponding to a period $T^* =
0.08912199969159$ s. The fixed point associated to the bifurcating
orbit is
$x^*=\left[\begin{array}{cc}5.09700788184250&0\end{array}\right]'.$
These values were detected firstly numerically and then obtained
analytically by solving \eqref{eq:fixedp} through an algebraic
manipulation software (For the sake of brevity we leave out the
computer algebra here).
\subsection{A locally piecewise-linear continuous map}
Let $\delta x_n$ and $\delta T$ be sufficiently small variations
of the state and parameter from the bifurcation point $x^*, T^*$.
We can then linearize the map $x_{n+1}=P(x_n,T)$ in
\eqref{eq:lmap} about this point as:
\begin{eqnarray}
\delta x_{n+1}=\frac{\partial P({x}^{*},T^*)}{\partial x_n} \delta
x_n + \frac{\partial P({x}^{*},T^*) }{\partial T}\delta T.
\label{eq:linearmap}
\end{eqnarray}
For the computation of $\frac{\partial P}{\partial x_n}$ it is
essential to take into account the implicit dependance of $t_i$ on
$x_n$ and $T$. Hence, using implicit differentiation, we have
\begin{eqnarray}
\frac{\partial P({x}_n,T)}{\partial x_n}&=&\frac{\partial
P({x}_n)}{\partial x_n}+\frac{\partial P(t_i)}{\partial
t_i}\frac{\partial t_i(x_n)}{\partial
 x_n}. \label{eq:dPdx}
\end{eqnarray}

Using \eqref{eq:lmap}, we can then write
\begin{eqnarray}
\frac{\partial P(x_n)}{\partial
x_n}&=&\phi_{\frac{T}{2}}(I+\phi_{-t_i}R\phi_{t_i}){\phi_{\frac{T}{2}}}
\label{Eq:dP_dxn}\\
\frac{\partial P(t_i)}{\partial
t_i}&=&\phi_{\frac{T}{2}}\left({\phi}_{t_i}'R\phi_{-t_i}-\left(\phi_{t_i}R{\phi}_{-t_i}'\right)\phi_{\frac{T}{2}}{x}_n-{\phi}_{t_i}'R
y_{-t_i}+\phi_{t_i}R {y}_{-t_i}'\right). \label{Eq:dP_dti}
\end{eqnarray}

Moreover, using implicit differentiation theorem, from
\eqref{eq:H} we have:
\begin{eqnarray}
\nonumber \frac{\partial H(x_n,t_i(x_n))}{\partial
x_n}=\frac{\partial H(x_n)}{\partial x_n}+\frac{\partial
H(t_i)}{\partial t_i}\frac{\partial t_i(x_n)}{\partial x_n}=0.
\end{eqnarray}
The above expression can be used to compute the remaining term in
\eqref{eq:dPdx} as:
\begin{eqnarray}
\frac{\partial t_i(x_n)}{\partial x_n}=-\left(\frac{\partial
H(t_i)}{\partial t_i}\right)^{-1} \frac{\partial H(x_n)}{\partial
x_n}, \label{Eq:dti_dxn}
\end{eqnarray}
where
\begin{eqnarray}
\nonumber \frac{\partial H(t_i)}{\partial t_i}&=&-h \cdot
\left(\phi_{\frac{T}{2}-t_i}' x_n - y_{-t_i}'\right),\\
\nonumber \frac{\partial H(x_n)}{\partial x_n}&=&h \cdot
\phi_{\frac{T}{2}-t_i},
\end{eqnarray}
and ${h}=\left[\begin{array}{cc}1 &0\end{array}\right]$.

After substituting \eqref{Eq:dP_dxn},\eqref{Eq:dP_dti} and
\eqref{Eq:dti_dxn} in \eqref{eq:dPdx} we obtain
\begin{multline}
\left.\frac{\partial P(x_n,T)}{\partial x_n}\right|_{x_n=x^* \atop
T=T^*}=\\
\phi_{\frac{T}{2}^*}\left((I+R)
 +\left(\left(R{\phi}_{0}'-{\phi}_{0}'R\right)\phi_{\frac{T}{2}^*}{x}^*+{\phi}_{0}'R
 y_{0}-R {y}_{0}'\right)
 \frac{h}{h \cdot
\left(\phi_{\frac{T}{2}^*}' x^* -
y_{0}'\right)}\right){\phi_{\frac{T}{2}^*}}.\\
\label{eq:ftaylormapcornerX} \end{multline}

In an analogous way, for the computation of $\frac{\partial
P}{\partial T}$, it is essential to take into account the implicit
dependance of $t_i$ on $x_n$ and $T$. Hence, by using implicit
differentiation, we have
\begin{eqnarray}
\frac{\partial P({x}_n,T)}{\partial T}&=&\frac{\partial
P(T)}{\partial T}+\frac{\partial P(t_i)}{\partial
t_i}\frac{\partial t_i(T)}{\partial
 T}. \label{eq:dPdT}
\end{eqnarray}

Using \eqref{eq:lmap}, we can then write
\begin{multline}
\frac{\partial P(T)}{\partial
T}=\\\left(\phi_T'+\frac{1}{2}\phi_{\frac{T}{2}+t_i}'R\phi_{\frac{T}{2}-t_i}+\frac{1}{2}\phi_{\frac{T}{2}+t_i}R\phi_{\frac{T}{2}-t_i}'\right)x_n
 -\frac{1}{2}\phi_{\frac{T}{2}+t_i}'R y_{-t_i}-\phi_{\frac{T}{2}+t_i}R \frac{\partial y_{-t_i,T}}{\partial
 T}.
 \label{Eq:dP_dT}
\end{multline}

Again, from \eqref{eq:H} we have:
\begin{eqnarray}
\nonumber \frac{\partial H(x_n,t_i(x_n))}{\partial
T}=\frac{\partial H(T)}{\partial T}+\frac{\partial
H(t_i)}{\partial t_i}\frac{\partial t_i(T)}{\partial T}=0,
\end{eqnarray}
that can be used to compute the remaining term in \eqref{eq:dPdT}.
Namely, we obtain:
\begin{eqnarray}
\frac{\partial t_i(T)}{\partial T}=-\left(\frac{\partial
H(t_i)}{\partial t_i}\right)^{-1} \frac{\partial H(T)}{\partial
T}, \label{Eq:dti_dT}
\end{eqnarray}
where
\begin{eqnarray}
\nonumber \frac{\partial H(t_i)}{\partial t_i}&=&-h \cdot
\left(\phi_{\frac{T}{2}-t_i}' x_n - y_{-t_i}'\right)\\
\nonumber \frac{\partial H(T)}{\partial T}&=&h \cdot\left(
\frac{1}{2}\phi_{\frac{T}{2}-t_i}' x_n-\frac{\partial
y_{-t_i,T}}{\partial T}\right)
\end{eqnarray}
and
$$\frac{\partial y_{t,T}}{\partial
T}=\left[\begin{array}{c}-\frac{t}{T}c'(t)\\-\frac{1}{T}c'(t)-\frac{t}{T}c''(t)\end{array}\right].$$
Finally, substituting \eqref{Eq:dP_dti}, \eqref{Eq:dP_dT} and
\eqref{Eq:dti_dT} into \eqref{eq:dPdT}, yields
\begin{multline}
\left.\frac{\partial P(x_n,T)}{\partial T}\right|_{x_n=x^* \atop
T=T^* }=\\
\left(\phi_{T^*}'+\frac{1}{2}\phi_{\frac{T}{2}^*}'R\phi_{\frac{T}{2}^*}+\frac{1}{2}\phi_{\frac{T}{2}^*}R\phi_{\frac{T}{2}^*}'\right)x^*
 -\frac{1}{2}\phi_{\frac{T}{2}^*}'R y_{0}-\phi_{\frac{T^*}{2}}R \frac{\partial y_{0,T^*}}{\partial T}
  \\
+\phi_{\frac{T}{2}^*}\left(\left(R{\phi}_{0}'-{\phi}_{0}'R\right)\phi_{\frac{T}{2}^*}x^*+{\phi}_{0}'R
y_{0}-R {y}_{0}'\right)\cdot\frac{h \cdot \left(
\frac{1}{2}\phi_{\frac{T}{2}^*}' x^*-\frac{\partial
y_{0,T^*}}{\partial T}\right)}{h \cdot \left(\phi_{\frac{T}{2}^*}'
x^* - y_{0}'\right)}. \label{eq:ftaylormapcornerT}
\end{multline}

We can then compute explicitly these quantities for the
cam-follower system of interest. In particular, after some
algebraic manipulation, we have:
\begin{equation}
A:=\frac{\partial P}{\partial x_n}(x^*,T^*)=\phi_{\frac{T}{2}^*}
\left[\begin{array}{cc}-r& 0\\-\frac{(1+r)(2\zeta {c}_0'+{
c}_0''+{\omega}^{2}_0{q}^{*}_d)}{{q}^{,*}_{d}-{
c}_0'}&-r\end{array}\right]\phi_{\frac{T}{2}^*} \label{eq:Px}
\end{equation}

and
\begin{multline}
B:=\frac{\partial P}{\partial T}(x^*,T^*)=
\frac{1}{2}\phi_{\frac{T}{2}^*}\left[\begin{array}{c}{q}^{*}_{d}\\-r{q}^{,*}_{d}+(1+r)c_0'\end{array}\right]+
\frac{1}{2}\phi_{\frac{T}{2}^*}'\left[\begin{array}{c}{{q}^{,*}_{d}}\\-{r}{q}^{,,*}_{d}-\frac{2(1+r)}{T^*}c_0'\end{array}\right]\\
+\frac{1}{2}\phi_{\frac{T}{2}^*}\frac{(1+r){q}^{,*}_d}{{q}^{,*}_{d}-{
c}_0'}\left[\begin{array}{c}{q}^{,*}_{d}-c_0'\\2\zeta {c}_0'+{
c}_0''+{\omega}^{2}_0{q}^{*}_d\end{array}\right].
\label{eq:PT}
\end{multline}

Note that both the matrices $A$ and $B$ as defined by
\eqref{eq:Px}-\eqref{eq:PT} depend on the value of the second
derivative of the cam acceleration $c_0''$ at the impact point.
Therefore the map is actually piecewise-linear locally to the
bifurcation point where the cam acceleration is discontinuous,
i.e.
$$
c_{0}''^-:=\lim_{t \rightarrow 0^-}{ c''(t)} \neq \lim_{t
\rightarrow 0^-}{ c''(t)} := c_0''^+.
$$

Then, the local map can be expressed as:
\begin{eqnarray}
\label{eq:pwlf}\delta x_{n+1}=\left\{\begin{array}{cc}A^- \delta
x_n +B^{-}\delta T,
&\text{If}\hspace{2mm} C \cdot \delta x_n + D  \cdot \delta T < 0,\\
&
\\
A^{+}\delta x_n +B^{+}\delta T, &\text{If}\hspace{2mm} C \cdot
\delta x_n + D  \cdot  \delta T>0,\end{array}\right.
\end{eqnarray}
where
$$
A^\pm = \frac{\partial P^\pm}{\partial x}, \quad B^\pm =
\frac{\partial P^\pm}{\partial T},
$$
with the index $\pm$ indicating whether the matrices are evaluated
with ${c}_0''={{c}_0''^-}$ or ${c}_0''={{c}_0''^+}$.

We have established that close to the corner-impact bifurcation
point, the dynamics of the follower can be studied by means of the
local mapping \eqref{eq:pwlf}.

Now, from \eqref{eq:lmap}, the global Poincar\'e map is known to
be a continuous function of the cam position and velocity through
the term $y_{-t_i}$. Moreover, the map is independent from the cam
acceleration. It follows, that the map is continuous at the
bifurcation point, \textit{i.e.} we must have that
$$A^- \delta x_n + B^- \delta T =A^+ \delta x_n + B^+ \delta T,$$
when
$$C \delta x_n + D \delta T=0.$$
Therefore we have
$$C=h\cdot(A^+ - A^-), \hspace{5mm}\text{and}\hspace{5mm}D=h\cdot(B^+ - B^-).$$

Substituting the numerical values of the map parameters for the
cam follower system of interest, we obtain the following
analytical estimates of the map matrices:

\begin{eqnarray}
 \nonumber {A}^-=\left[\begin{array}{cc}0.82093496821478
 &0.01346530915655\\
   2.52012201452530 &  0.82093496821478\end{array}\right],\
 {B}^-=\left[\begin{array}{c} -51.62757990297\\
  -5455.79455977621\end{array}\right],
\end{eqnarray}
\begin{eqnarray}
 \nonumber {A}^+ = \left[\begin{array}{cc}0.68571072072040
 &-0.07351052377964\\
   2.30988433707948  & 0.68571072072040\end{array}\right],\
 {B}^+ = \left[\begin{array}{c}  208.11740649865\\
  -5051.96030903248\end{array}\right]
\end{eqnarray}
and
$$
C=\left[-0.13522424749438\ -0.08697583293619\right],\ D=
259.7449864016200.
$$

\subsubsection{Numerical Validation}
We will now validate our numerical findings by comparing the map
\eqref{eq:pwlf}, which was derived analytically, with the
numerical estimates of the mapping obtained by means of simulation
and an optimized fitting algorithm close to the bifurcation point.

To derive such an estimate, we use an accurate event-driven
numerical algorithm to simulate the cam dynamics over one period
starting from a set of $M$ different initial conditions and
parameter values. Namely, say $\delta \bar x_n$ the vector of $M$
possible perturbations of $x^*$ and $\delta \bar T$ the vector of
$M$ possible perturbations of $T$. We then simulate the cam
dynamics from each of the perturbed initial conditions and
parameter values to obtain the vector $\delta \bar
x_{n+1}=x^*-x_{n+1}$ after one period. We repeat the set of
simulation twice, once with the cam acceleration set to $c_0''^+$
and once with the acceleration set to $c_0''^-$. In so doing, we
obtain numerically the vectors
$$\delta \bar {x}^{\pm}_{n+1}=\left[\begin{array}{ccccc}
\delta \bar {x}^{1}_{n+1}&\dots&\delta \bar
{x}^{m}_{n+1}&\dots&\delta \bar {x}^{M}_{n+1}\end{array}\right].$$

We then use a least-squares fitting algorithm to estimate the
matrices $\hat A^\pm$ and $\hat B^\pm$ that minimize the error
$$e=\left\| \delta \bar { x}^{\pm}_{n+1}-\left[\begin{array}{ccc}{\hat A}^{\pm}
&|&{\hat B}^{\pm}\end{array}\right]\left[\begin{array}{c}\delta \bar {x}_{n}\\
\delta \bar T\end{array}\right]\right\|^2.$$

The estimated map matrices found using this numerical strategy are
\begin{eqnarray}
 \nonumber \hat{A}^-=\left[\begin{array}{cc}0.82093497830369   &0.01346530945739\\
   2.52012201542191   &0.82093496286678\end{array}\right],
 \hat{B}^-=\left[\begin{array}{c} -51.62757113994\\
  -5455.79411324739\end{array}\right],
\end{eqnarray}
\begin{eqnarray}
 \nonumber \hat{A}^+ = \left[\begin{array}{cc}0.68571065978423&-0.07351053029558\\
   2.30988432418263&   0.68571073479454\end{array}\right],
 \hat{B}^+ = \left[\begin{array}{c} 208.11731732063\\
  -5051.95951604729\end{array}\right].
\end{eqnarray}

We notice that these numerical estimates are almost identical (up
to at least 5 decimal places) to those obtained analytically
earlier in the paper. This validates our analysis and shows the
reliability of the analytical derivation used to get a leading
order estimate of the Poincar\'e map close to the bifurcation
point under investigation.

\subsection{Classification of the Non-Smooth Bifurcation Scenario}

We can now use the locally derived map (analytical or numerical)
to classify and explain the bifurcation scenario due to the
corner-impact bifurcation detected in the cam-follower system of
interest. In particular, the map derived above is a piecewise
linear continuous map. As the cam rotational speed is increased,
the period $T$ of the forcing provided by the cam varies.
Correspondingly, at the corner-impact bifurcation point ($\delta T
=0$), the map fixed point undergoes a border collision.  Feigin
strategy for border-collision bifurcations can then be used to
classify the corner-impact bifurcation scenario \cite{BeFe:99}.

The idea is to start by recasting the map \eqref{eq:pwlf} into a
canonical form following the procedure presented in
\cite{diBernardo:06}. Specifically,
\begin{enumerate}
\item We eliminate the term depending on $\delta T$
by considering an appropriate change of coordinates. In particular
if we say $c_1$ and $c_2$ the coefficients of $C$, we choose:
\begin{eqnarray}
 \nonumber \delta {\tilde x}_n^1 &=& \delta x_n^1+D\frac{\mu}{c_1},\\
 \nonumber \delta {\tilde x}_n^2 &=& \delta x^2_n,
\end{eqnarray}
so that the map becomes
\begin{eqnarray}
\nonumber \delta \tilde{x}_{n+1}=\left\{\begin{array}{cc}A^-
\delta \tilde  x_n + \tilde B \delta T,
&\text{If}\hspace{2mm} C \cdot \delta \tilde x_n   < 0,\\
&
\\
A^{+}\delta \tilde x_n +\tilde B \delta T, &\text{If}\hspace{2mm}
C \cdot \delta \tilde x_n >0,\end{array}\right.
\end{eqnarray}
where
$$\tilde
B=\left[\begin{array}{c}{b}^{-}_1-\frac{{a}^{-}_{11}}{c_1}d\\{b}^{-}_2-\frac{{a}^{-}_{21}}{c_1}d\end{array}\right]
=\left[\begin{array}{c}{b}^{+}_1-\frac{{a}^{+}_{11}}{c_1}d\\{b}^{+}_2-\frac{{a}^{+}_{21}}{c_1}d\end{array}\right]
=\left[\begin{array}{c}1525.26226128059\\
  -615.02768162765\end{array}\right],$$
with $a^\pm_{ij}$ being the coefficients of $A^\pm$.
\item Then, using the strategy presented in \cite{diBernardo:06,diBnf:03}, we
consider the change of coordinates $x=W^{-1}{\tilde x}$ where the
matrix $W$ is obtained as $W=T^-O^-$ with
  $$O^-=\left[\begin{array}{c}C\\
  C A^-\end{array}\right], T^-=\left[\begin{array}{cc}1&0\\
  {d}^{-}_{1}&1\end{array}\right],  $$
where $d^-_1$ is the linear coefficient of the characteristic
polynomial of $A^-$ given by
$p^-(\lambda)={\lambda}^{2}+{d}^{-}_{1}\lambda+{d}^{-}_{2}$.
Applying such a similarity transformation, the map matrices
become:
 $$
 \bar A^-=\left[\begin{array}{cc}1.64186993642956&1\\-0.64&0\end{array}\right],\
 \bar A^+=\left[\begin{array}{cc}1.37142144144080&1\\-0.64&0\end{array}\right],$$
 and
 $$
 \bar B=
 \left[\begin{array}{c}152.75990\\207,79599\end{array}\right],
 \hspace{1cm}\bar C=\left[\begin{array}{cc}1& 0\end{array}\right].$$
\end{enumerate}

\begin{figure}
\begin{center}
\includegraphics[width=8cm]{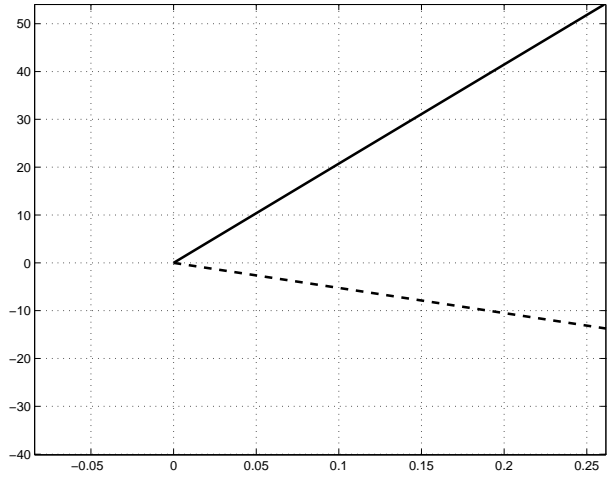}
\end{center}
\caption{Numerical bifurcation diagram of the local map
\eqref{eq:pwlf} with the analytically estimated matrices. The
border collision when $\delta T=0$ corresponds to the
corner-impact bifurcation point at $\omega \approx 673.2$ rpm.
Note that as predicted a nonsmooth fold scenario is observed with
no fixed point existing for $\delta T <0$ and two coexisting fixed
points, one stable, the other unstable for $\delta
T>0$.}\label{fig:bif_time2}
\end{figure}

As explained in \cite{BeFe:99,diBernardo:06}, we can now classify
the type of bifurcation scenario observed at the bifurcation point
under investigation by computing the map eigenvalues on both sides
of the boundary.  For the case under investigation, we have that:
(i) the eigenvalues of $A^-$ are $\lambda_1^-=1.0052$ and
$\lambda_2^-=0.6367$; (ii) the eigenvalues of $A^+$ are
$\lambda_{1,2}^+=0.6857 \pm \mathbf{j} 0.4120$. Hence, according
to Feigin's classification strategy, since the total number of
real eigenvalues greater than unity on both sides of the boundary
is odd, the bifurcating fixed point will undergo a nonsmooth
saddle node bifurcation and ceases to exist \cite{BeFe:99}. This
is in perfect agreement with what observed numerically as shown in
Fig. \ref{fig:bif_time2}, where the local bifurcation scenario
observed in the map is shown.

Therefore, we can explain the sudden transition to chaos observed
in the cam-follower system under investigation as due to the
occurrence of a corner-impact bifurcation. Namely, the
corner-impact is associated to a nonsmooth-fold scenario causing
the disappearance of the stable impacting solution undergoing the
bifurcation. This causes trajectories to leave the local
neighborhood where they are confined before the bifurcation and
converge  towards the stable coexisting chaotic attractor when
$\omega$ is decreased below the corner-impact bifurcation point.

Hence, we can conclude that corner-impact bifurcations in cam
follower systems can indeed lead to dramatic changes of the system
qualitative behavior including sudden transitions from periodic
solutions to chaos.

\section{Conclusions}
We have studied a novel type of discontinuity-induced bifurcation
in a class of mechanical devices widely used in applications:
cam-follower systems. Using a representative second-order model of
the follower, we have shown that its dynamics can undergo several
bifurcations including sudden transitions to chaos as the cam
rotational speed is varied. We analysed in detail the
corner-impact bifurcation of a one-periodic solution characterised
by one impact per period. In particular, we observed that the
system behaviour undergoes dramatic changes when the impact occurs
at a point where the cam profile is discontinuous. Using the
concept of discontinuity mappings, we derived analytically the
Poincar\'e map associated to  the bifurcating orbit in the case
where the cam profile has a discontinuous acceleration. Then,
using the classification strategy for border-collision
bifurcations, we proved that the corner-impact causes the fixed
point associated to the bifurcating orbit to undergo a nonsmooth
saddle-node bifurcation. Namely, the fixed point ceases to exist,
with the trajectories being attracted towards a chaotic invariant
set.

We wish to emphasize that:
\begin{itemize}
\item the analysis presented above applies with minor changes to
the case of impact oscillators forced by signals with
discontinuous second derivative. As shown above, this leads to
maps which are locally piecewise linear continuous close to a
corner-impact bifurcation point. This extends the analysis
presented in \cite{BuPi:06} for the case of an impact oscillator
forced by a function with discontinuous first derivative. We
conjecture that the properties of the local mapping depend on the
degree of discontinuity of the forcing signal. This is the subject
of ongoing work.

\item As shown in \cite{BeBu:01}, discontinuity-induced
bifurcations in flows are usually associated to maps which are not
piecewise linear. Grazing bifurcations of limit cycles where known
to be associated to maps with square-root singularities in
impacting systems and Filippov systems \cite{diBernardo:01b} or
maps with higher order nonlinear terms in the case of
piecewise-smooth continuous flows \linebreak (PWSC). The only
cases in the literature where the map was indeed found to be
piecewise linear-continuous were corner-collisions in PWSC systems
and grazing sliding bifurcations in Filippov systems. So far, no
evidence was given of a bifurcation event in impacting systems
associated to locally piecewise-linear continuous maps. The
corner-impact bifurcation scenario presented in this paper fills
this gap in the literature.

\item We believe cam follower systems are a particularly useful
set-up to show generically the behaviour of impacting systems with
discontinuous forcing.
\end{itemize}

Finally, the results presented here can open a way to future work
towards a better understanding of the complex dynamics of
cam-follower systems. This can lead to less conservative solutions
to detachment avoidance, hopefully without recurring to highly
stiff closing springs and maybe  active control strategies.
\section*{Acknowledgments}
The authors wish to thank the anonymous reviewers whose comments
led to a consistent revision of the original version of this
manuscript. They also gratefully acknowledge support from the
European Union (EU Project SICONOS - V Framework Programme,
IST2001-37172) and the project MIUR-PRIN MACSI funded by the
Italian Ministry for Research and University. The paper was
completed during a research visit of the authors at the Centre de
Recerca Matematica in Barcelona thanks to support from the
Government of Catalunya.
\section*{Appendix A - Cam Profile}
\begin{figure}[bh]
\setlength{\unitlength}{1mm}
\begin{picture}(65,80)(0,0)
\begin{texdraw}
\drawdim cm \linewd 0.02 \arrowheadtype t:F \arrowheadsize l:0.16
w:0.08
 \move(3.25 4.0) 
 \setgray 0.5 \lpatt() \rmove(-3 0) \ravec(6 0)
 \lpatt() \move(3.25 0.5) \ravec(0 7)
 \linewd 0.03
 \setgray 0
 \move(3.25 4.0) \larc r:2 sd:20 ed:160 
 \move(3.25 4.0) \larc r:3.0716 sd:250 ed:290 
 \move(4.6886 3.4764 ) \larc r:3.5310 sd:160 ed:216.5976 
 \move(1.8114 3.4764 ) \larc r:3.5310 sd:-36.5976 ed:20 
 \move(2.4560 1.8184) \larc r:0.745 sd:216.5976 ed:250 
 \move(4.0440 1.8184) \larc r:0.745 sd:290 ed:-36.5976 
 \linewd 0.02
 \setgray 0.6
 \move(1.8114 3.4764 ) \lvec(4.6462    1.3712) 
 \move(1.8114 3.4764 ) \lvec(5.1294 4.6840) 
 \move(3.25 4.0) \lvec(4.3006    1.1136) 
 \lpatt(0.1 0.07)
 \move(3.25 4.0) \lvec(4.6462    1.3712) 
 \lpatt()
 \setgray 0
 \move(3.25 4.0) \fcir f:0 r:0.05 
 \move(4.6886 3.4764 ) \fcir f:0 r:0.05 
 \move(1.8114 3.4764 ) \fcir f:0 r:0.05 
 \move(2.4560 1.8184) \fcir f:0 r:0.05 
 \move(4.0440 1.8184) \fcir f:0 r:0.05 
 \move(1.3706 4.6840) \fcir f:0.6 r:0.05 
 \move(5.1294 4.6840) \fcir f:0.6 r:0.05 
 \move(1.1967    4.0000) \fcir f:0.6 r:0.05 
 \move(5.3033    4.0000) \fcir f:0.6 r:0.05 
 \move(1.8538    1.3712)  \fcir f:0.6 r:0.05 
 \move(4.6462    1.3712)  \fcir f:0.6 r:0.05 
 \move(2.1994    1.1136)  \fcir f:0.6 r:0.05 
 \move(4.3006    1.1136)  \fcir f:0.6 r:0.05 
 \move(3.25    6)  \fcir f:0.6 r:0.05 
 \textref h:R v:B   \htext(3.2 4.1){\textit{O}}
 \textref h:R v:T   \htext(1.8114 3.4764){\textit{$O_1$}}
 \textref h:R v:T   \htext(4.0440 1.8184){\textit{$O_2$}}
 \textref h:L v:T   \htext(4.3006    1.0136){\textit{Q}}
 \textref h:L v:T   \htext(4.6462    1.2712){\textit{P}}
 \textref h:L v:B   \htext(5.1794  4.7840){\textit{R}}
 \textref h:L v:B   \htext(5.3533    4.1000){\textit{S}}
 \textref h:C v:T   \htext(6  3.9){\textit{x}}
 \textref h:R v:C   \htext(3.2    7.1){\textit{y}}
 \textref h:L v:B   \htext(3.32    6.05){\text{$c(\theta)$}}
 \textref h:C v:c   \htext(3.25 -0.1){\textit{(a)}}
\end{texdraw}
\end{picture}
\begin{picture}(65,80)(0,0)
\begin{texdraw}
\drawdim cm \linewd 0.02 \arrowheadtype t:F \arrowheadsize l:0.16
w:0.08
\move(3.25 4.0) 
 \setgray 0.5 \lpatt() \rmove(-3 0) \ravec(6 0)
 \lpatt() \move(3.25 0.5) \ravec(0 7)
 \linewd 0.03
 \setgray 0
 \move(3.25 4.0) \larc r:2 sd:200 ed:340 
 \move(3.25 4.0) \larc r:3.0716 sd:70 ed:110 
 \move(4.6886 4.5236 ) \larc r:3.5310 sd:143.4024 ed:200 
 \move(1.8114 4.5236 ) \larc r:3.5310 sd:-20 ed:36.5976 
 \move(2.4560 6.1816) \larc r:0.745 sd:110 ed:147.1976 
 \move(4.0440 6.1816) \larc r:0.745 sd: 32.8367 ed:70 
 \linewd 0.02
 \setgray 0.6
 \move(4.6886 4.5236 ) \lvec(1.8538    6.6288) 
 \move(4.6886 4.5236 ) \lvec(1.3706 3.3160) 
 \move(3.25 4.0) \lvec(2.1994    6.8864) 
 \lpatt(0.1 0.07)
 \move(3.25 4.0) \lvec(1.8538    6.6288) 
 \lpatt()
 \setgray 0
 \move(3.25 4.0) \fcir f:0 r:0.05 
 \move(4.6886 4.5236 ) \fcir f:0 r:0.05 
 \move(1.8114 4.5236 ) \fcir f:0 r:0.05 
 \move(2.4560 6.1816) \fcir f:0 r:0.05 
 \move(4.0440 6.1816) \fcir f:0 r:0.05 
 \move(1.3706 3.3160) \fcir f:0.6 r:0.05 
 \move(5.1294 3.3160) \fcir f:0.6 r:0.05 
 \move(1.1967    4.0000) \fcir f:0.6 r:0.05 
 \move(5.3033    4.0000) \fcir f:0.6 r:0.05 
 \move(1.8538    6.6288)  \fcir f:0.6 r:0.05 
 \move(4.6462    6.6288)  \fcir f:0.6 r:0.05 
 \move(2.1994    6.8864)  \fcir f:0.6 r:0.05 
 \move(4.3006    6.8864)  \fcir f:0.6 r:0.05 
 \textref h:C v:c   \htext(3.25 -0.1){\textit{(b)}}
\end{texdraw}
\end{picture}
  \caption{Cam profile definition. \textit{(a)} $\theta=0$. \textit{(b)} $\theta=\pi$. .}
  \label{Fig:camprofileeq}
\end{figure}
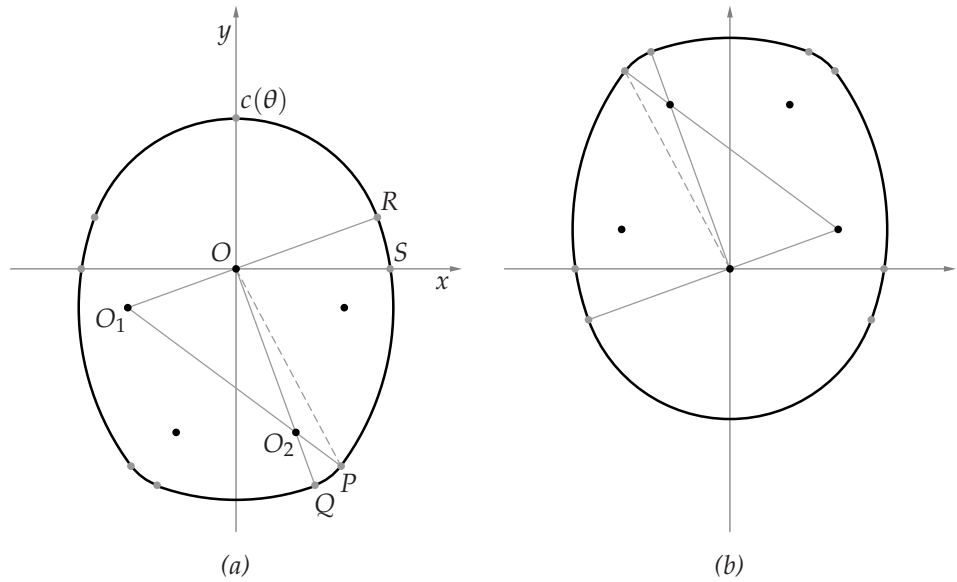

We report below the analytical description of the representative
cam profile considered in this paper. As shown in Fig.
\ref{Fig:camprofileeq}, in this case the cam profile is the result
of a geometrical based design.

The lift profile $c(\theta)$ can be defined from the construction
as a piecewise smooth function of the angle $\theta$ as:
\begin{eqnarray}
c(\theta) &=& \left\{\begin{array}{clc}
   c_0(\theta)  &   \text{ If } 0< \theta \leq \frac{\pi}{2} - \theta_1,\\
   c_1(\theta)  &   \text{ If } \frac{\pi}{2} - \theta_1< \theta \leq \frac{\pi}{2} - \theta_2,\\
   c_2(\theta)  &   \text{ If } \frac{\pi}{2} - \theta_2< \theta \leq \frac{\pi}{2} - \theta_3,\\
   c_3(\theta)  &   \text{ If } \frac{\pi}{2} - \theta_3< \theta \leq
   \pi,
\end{array}\right.
\end{eqnarray}
\begin{eqnarray}
\nonumber
 c_0(\theta)&=&\rho_0\\
\nonumber
 c_1(\theta)&=&-\kappa_1\sin({\theta+\theta_1})+({\rho^2_1}-{{\kappa^2_1}}{\cos({\theta+\theta_1}})^2)^{\frac{1}{2}}\\
\nonumber
 c_2(\theta)&=&\kappa_2 \sin({\theta+\theta_3})+({\rho^2_2}-{{\kappa^2_2}}{\cos({\theta+\theta_3}})^2)^{\frac{1}{2}}\\
\nonumber
 c_3(\theta)&=&\rho_3
\end{eqnarray}
where $\theta_1=\angle SOR $, $\theta_2=\angle SOP $,
$\theta_3=\angle
 SOQ$. Additionally,  $\kappa_i$ and $\rho_i$ are constant parameter given by our
particular geometrical construction of the cam as ({\em See }
Fig.\ref{Fig:camprofileeq})
\begin{eqnarray}
\begin{array}{lll}
\kappa_1=\|\overline{OO_1}\|,&\rho_0=\|\overline{OR}\|,&\rho_2=\|\overline{O_2P}\|,\\
\kappa_2=\|\overline{OO_2}\|,&\rho_1=\|\overline{O_1R}\|,&\rho_3=\|\overline{OQ}\|.\\
\end{array}
\end{eqnarray}

\bibliographystyle{plain}

\end{document}